\documentclass[fleqn,3p,onecolumn,11pt,nofootinbib,showkeys,showpacs]{revtex4-1}

\usepackage[small,justification=centerlast]{caption}
\usepackage{graphicx}
\usepackage{bm}
\usepackage{mathrsfs}
\usepackage[latin1]{inputenc}
\usepackage{array}
\usepackage{psfrag}
\usepackage{dsfont}
\usepackage{epsfig}
\usepackage{titletoc}
\usepackage{float}   
\usepackage{wrapfig} 
\usepackage{amsmath}\usepackage{amssymb,stmaryrd}
\usepackage{array}
\usepackage{epsfig}
\usepackage{cancel}
\usepackage[dvips]{feynmp}
\usepackage{upgreek}
\usepackage{subfig}
\usepackage{tabularx}
\usepackage{xcolor}
\usepackage{units}
\usepackage{textcomp}
\usepackage{wasysym}
\usepackage{dutchcal}
\usepackage{amsthm}
\newtheorem{conjecture}{Proposition}

\DeclareMathAlphabet{\standardcal}{OMS}{zplm}{m}{n}

\usepackage{textfit} 

\usepackage{hyperref}

\renewcommand{\eqref}[1]{\mbox{Eq.~(\ref{#1})}}
\newcommand{\tabref}[1]{\mbox{Tab.~\ref{#1}}}
\newcommand{\figref}[1]{\mbox{Fig.~\ref{#1}}}
\newcommand{\secref}[1]{\mbox{Sec.~\ref{#1}}}

\newcommand{\appref}[1]{\mbox{App.~\ref{#1}}}

\begin{document}

\title{From classical Lagrangians to Hamilton operators \\ in the Standard-Model Extension}

\author{M. Schreck} \email{marco.schreck@ufma.br}
\affiliation{Departamento de F\'{i}sica, Universidade Federal do Maranh\~{a}o \\
65080-805, S\~{a}o Lu\'{i}s, Maranh\~{a}o, Brazil
}

\begin{abstract}
In this article we investigate whether a theory based on a classical Lagrangian for the minimal Standard-Model Extension (SME) can be
quantized such that the result is equal to the corresponding low-energy Hamilton operator obtained from the field-theory description. This
analysis is carried out for the whole collection of minimal Lagrangians found in the literature. The upshot is that first quantization
can be performed consistently. The unexpected observation is made that at first order in Lorentz violation and at second order in the
velocity the Lagrangians are related to the Hamilton functions by a simple transformation. Under mild assumptions, it is shown that this
holds universally. This result is used successfully to obtain classical Lagrangians for two complicated sectors of the minimal SME
that have not been considered in the literature so far. Therefore, it will not be an obstacle anymore to derive such Lagrangians even
for involved sets of coefficients --- at least to the level of approximation stated above.
\end{abstract}

\maketitle

\newpage
\setcounter{equation}{0}
\setcounter{section}{0}
\renewcommand{\theequation}{\arabic{section}.\arabic{equation}}

\section{Introduction}

Fundamental physics of the 21st century will be governed by the search for a theory of quantum gravity. This will ultimately bring the
field of {\em CPT}- and Lorentz violation more into the focus of high-energy physics. One of the basic and most essential results obtained
in this context is that Lorentz violation can arise naturally in closed-string field theory \cite{Kostelecky:1988zi,Kostelecky:1991ak,
Kostelecky:1994rn}. Besides that, a violation of Lorentz invariance was shown to occur in other realms of high-energy physics or alternative
approaches to quantum gravity such as loop quantum gravity \cite{Gambini:1998it,Bojowald:2004bb}, theories of noncommutative spacetimes
\cite{AmelinoCamelia:1999pm,Carroll:2001ws}, spacetime foam models \cite{Klinkhamer:2003ec,Bernadotte:2006ya,
Hossenfelder:2014hha}, quantum field theories in backgrounds with nontrivial topologies \cite{Klinkhamer:1998fa,Klinkhamer:1999zh}, and
last but not least, Ho\v{r}ava-Lifshitz gravity \cite{Horava:2009uw}.

These results have far-reaching consequences. First, they show that violations of the aforementioned fundamental symmetries may be signals
of physics at the Planck scale. Second, Lorentz invariance is the very base of the established theories, i.e., the Standard Model and
General Relativity. Hence, even if quantum gravity looks totally different from what physicists currently imagine these theories should be
subject to precise experimental tests. This is only possible based on a framework that extends both the Standard Model and General Relativity
to tell the experimentalists what effects they may expect to measure. The most general
framework is the Standard-Model Extension (SME) \cite{Colladay:1998fq}. The latter is an effective field theory based on the Standard Model,
which comprises all Lorentz-violating operators of mass dimension 3 and 4 that can be added to the Lagrange density. The nonminimal SME
\cite{Kostelecky:2009zp,Kostelecky:2011gq,Kostelecky:2013rta} additionally includes all higher-dimensional operators.
Whenever Lorentz violation is studied, {\em CPT} violation is automatically taken into account due to a theorem by Greenberg
\cite{Greenberg:2002uu}, which says that in effective field theory {\em CPT}-violation implies Lorentz noninvariance.

Various theoretical investigations of the SME have been carried out at tree-level \cite{Kostelecky:2000mm,oai:arXiv.org:hep-ph/0101087,
Casana:2009xs,Casana:2010nd,Klinkhamer:2010zs,Schreck:2011ai,Casana:2011fe,Hohensee:2012dt,Cambiaso:2012vb,Schreck:2013gma,Schreck:2013kja,
Schreck:2014qka,Colladay:2014dua,Casana:2014cqa,Albayrak:2015ewa} and higher-orders in perturbation theory \cite{Jackiw:1999yp,Chung:1999pt,
PerezVictoria:1999uh,PerezVictoria:2001ej,Kostelecky:2001jc,Altschul:2003ce,Altschul:2004gs,Colladay:2006rk,Colladay:2007aj,Colladay:2009rb,
Gomes:2009ch,Ferrero:2011yu,Casana:2013nfx,Scarpelli:2013eya,Cambiaso:2014eba,Santos:2014lfa,Santos:2015koa,Borges:2016uwl,Belich:2016pzc}.
These results are important as they demonstrate that the SME is a viable framework to investigate Lorentz violation. Therefore, experimental
studies are warranted as well where a great deal of sharp experimental bounds on the minimal SME already exist opening the pathway to covering
the leading-order operators of the nonminimal SME. A yearly updated compilation of all constraints can be found in \cite{Kostelecky:2008ts}.

After setting up the particle-physics part of the minimal SME in \cite{Colladay:1998fq}
the gravity part was established in \cite{Kostelecky:2003fs}. One of the most crucial results of the latter paper is that explicit
Lorentz violation is incompatible with gravity where the incompatibility is due to the Bianchi identities of Riemannian geometry.
Therefore, in a curved background Lorentz violation can only be studied consistently when the symmetry is broken spontaneously,
cf.~\cite{Kostelecky:1989jp,Kostelecky:1989jw,Bluhm:2008yt,Hernaski:2014jsa,Bluhm:2014oua,Maluf:2014dpa}. An alternative possibility of
circumventing the incompatibility could be to use a geometric concept other than Riemannian geometry. Because of that reason, Lorentz violation based
on Finsler geometry \cite{Finsler:1918,Cartan:1933,Bao:2000,Antonelli:1993} is currently being investigated extensively. Finsler geometry
can be regarded as Riemannian geometry without the quadratic restriction of line intervals, i.e., any possible interval obeying
certain reasonable properties can be considered. Line intervals can also involve preferred directions on the manifold, which is why
this geometric approach is interesting for people studying Lorentz violation.

To consider Finsler geometry in the SME a reasonable starting point is needed. General Relativity and possible extensions of it
reside in the realm of classical physics. However, the particle-physics part of the SME is a field theory concept, i.e., one has to
map the field theory description to a classical-physics analog. This was carried out for various cases of the minimal SME
\cite{Kostelecky:2010hs,Kostelecky:2011qz,Kostelecky:2012ac,Colladay:2012rv,Schreck:2014ama,Russell:2015gwa} and a number of
nonminimal cases \cite{Schreck:2014hga,Schreck:2015seb}. The results of these studies are Lagrangians for classical,
relativistic, pointlike particles including Lorentz violation based on the SME. Further analyses of these Lagrangians or
investigations in other sectors of the SME can be found in \cite{Silva:2013xba,Foster:2015yta,Colladay:2015wra,
Schreck:2015dsa,Silva:2015ptj}. The classical Lagrangians obtained were shown to be related to Finsler structures
\cite{Kostelecky:2011qz,Kostelecky:2012ac,Colladay:2012rv,Schreck:2014hga,Schreck:2015seb} and can possibly serve to
study explicit Lorentz violation in curved backgrounds, cf.~\cite{Kostelecky:2011qz,Schreck:2015dsa}.

The mapping investigated in the papers mentioned above starts at the quantum description of the SME and it ends at the classical
regime. Therefore, the motivation of the current article is to answer one question. Assuming that we have the classical Lagrangians
only and do not know about the field theory description of the SME, is it possible to quantize the classical theory to arrive at
the quantum-mechanical Hamiltonian based on the SME? Note that the SME is a relativistic field theory, which allows for obtaining
the corresponding low-energy Hamiltonian with the Foldy-Wouthuysen procedure \cite{Foldy:1949wa}. Hence, an alternative
method could be to expand the relativistic classical Lagrangians in the ratio of the velocity and the speed of light and to
perform quantization subsequently. Finding out whether or not this method works is the goal of the paper. In the course of
the analysis, the unexpected result is encountered that the classical Lagrangians and Hamilton functions considered are related by
a simple transformation at first order in Lorentz violation and at second order in the momenta. This observation may be
interesting and important in practice.

The paper is organized as follows. In \secref{sec:classical-hamilton-functions} we compile all Lagrangians obtained in the
literature so far. Thereby, the velocity-momentum correspondence and the classical Hamilton function is computed for each.
Section \ref{sec:first-quantization} is dedicated to the first quantization of the results. All classical momenta are
promoted to quantum operators and a suitable \textit{Ansatz} for a spin structure is introduced. It is shown that the
quantum-mechanical Hamilton operators can be obtained consistently. In \secref{sec:limit-of-lagrangians} the leading-order
expansion of each Lagrangian is investigated more closely. By doing so, we find the simple relation between the Lagrangians
and the Hamilton functions mentioned above. Under mild assumptions, it is shown that this result is valid in general. Subsequently,
we apply it to two complicated
cases of the minimal fermion sector not considered in the literature so far. Last but not least, all findings are discussed
and concluded on in \secref{sec:conclusions}. Throughout the paper, natural units with $\hbar=c=1$ are used unless otherwise
stated.

\section{Classical Hamilton functions}
\label{sec:classical-hamilton-functions}

In \cite{Kostelecky:2010hs} the procedure was set up to assign a classical Lagrangian to a particular case of the SME fermion sector.
Consider a quantum wave packet that is a superposition of plane-wave solutions to the free-field equations with a suitable smearing
function. If the smearing function in configuration space is chosen to fall off sufficiently fast outside of a localized region this
wave packet is interpreted as a particle in the classical limit. The physical propagation velocity of the packet corresponds to its
group velocity for most cases \cite{Brillouin:1960}. Denoting the four-momentum of a plane wave, which is part of the wave packet,
by $p_{\mu}$ and the four-velocity of the classical particle by $u^{\mu}$ we have the following five equations that govern the
mapping procedure:
\begin{subequations}
\label{eq:set-equations-lagrangians}
\begin{align}
\label{eq:dispersion-relation}
\mathcal{R}(p)&=0\,, \displaybreak[0]\\[2ex]
\label{eq:group-velocity-correspondence}
\frac{\mathrm{d}p_0}{\mathrm{d}p_i}&=-\frac{u^i}{u^0}\,,\quad i\in\{1,2,3\}\,, \displaybreak[0]\\[2ex]
\label{eq:euler-equation}
L&=-p_{\mu}u^{\mu}\,,\quad p_{\mu}=-\frac{\partial L}{\partial u^{\mu}}\,.
\end{align}
\end{subequations}
The first is the dispersion relation of the particular SME fermion sector considered. The second says that the group velocity of
the quantum wave packet shall correspond to the three-velocity of the classical particle. These are three equations, one for each
component. The last follows from the reasonable assumption of an action that is invariant under changes of parameterization.
For exhaustive discussions on that procedure we refer to~\cite{Kostelecky:2010hs}.

Within the current paper a classical Lagrangian $L=L(u^0,\mathbf{u})$ shall not be obtained but it is supposed to be the starting
point. The aim is to derive the quantum-mechanical Hamiltonian from this Lagrangian. To do so, the first step is to obtain
the particle energy as a function of velocity for which there are two possibilities. In proper-time parameterization, $u^0=1$ and
$\mathbf{u}=\mathbf{v}$ where $\mathbf{v}$ is the three-velocity
of the particle. According to $p_{\mu}=-\partial L/\partial u^{\mu}$ we obtain
\begin{equation}
\label{eq:energy-from-derivative-u0}
E=E(\mathbf{v})=\left.-\frac{\partial L}{\partial u^0}\right|_{\substack{u^0=1 \\ \mathbf{u}=\mathbf{v}}}\,.
\end{equation}
The second method is to derive the energy via a Legendre transformation, cf.~\eqref{eq:euler-equation}:
\begin{subequations}
\label{eq:legendre-transform}
\begin{align}
L&=-p_{\mu}u^{\mu}|_{\substack{u^0=1 \\ \mathbf{u}=\mathbf{v}}}=-E+\mathbf{p}\cdot\mathbf{v}\,, \\[2ex]
E&=\frac{\partial L}{\partial\mathbf{v}}\cdot\mathbf{v}-L\,,
\end{align}
\end{subequations}
where the spatial momentum $\mathbf{p}$ is understood to have upper indices. In what follows, both procedures are checked to lead to the same result,
which is the particle energy as a function of the three-velocity. For quantization, the Hamiltonian has to be computed from the
classical energy. To do so the energy is needed as a function of spatial momentum $\mathbf{p}$ instead of velocity $\mathbf{v}$.
Hence, it is necessary to solve
\begin{equation}
\mathbf{p}=\frac{\partial L}{\partial\mathbf{v}}\,,
\end{equation}
with respect to $\mathbf{v}$ to give replacement rules for the velocity in favor of the momentum. The result is the classical
Hamilton function $\mathcal{H}=\mathcal{H}(\mathbf{p})$, which forms the basis for quantization. In the forthcoming subsections,
this procedure will be carried out for all classical Lagrangians found for the minimal SME fermion sector. We will work at
first order in Lorentz violation and at second order in the momentum or velocity.

For each case of the SME fermion sector there are distinct Lagrangians for the particle and the antiparticle solutions. They
are related to each other by the replacement $m_{\psi}\mapsto -m_{\psi}$. In this article we will only consider the particle
Lagrangians as they deliver positive energies. Classically, there are no antiparticles after all. For the properties of the
SME fermion coefficients we refer to Table 1 in~\cite{Kostelecky:2013rta}.

\subsection{Operator \texorpdfstring{$\boldsymbol{\widehat{a}^{(3)}}$}{a3-hat}}

We start with the Lagrangian for the $a$ coefficients that are of mass dimension 1. It is based on the observer four-vector
$a^{(3)}_{\mu}=(a^{(3)}_0,\mathbf{a})_{\mu}$ with $\mathbf{a}\equiv (a^{(3)}_1,a^{(3)}_2,a^{(3)}_3)$. The Lagrange function can be
extracted from Eq.~(8) or Eq.~(12) in \cite{Kostelecky:2010hs} when setting all the other controlling coefficients to zero.
It is comprised of the standard square root term and an observer Lorentz scalar involving both the four-velocity and the observer
four-vector $a^{(3)}_{\mu}$:
\begin{equation}
\label{eq:lagrangian-a3}
L^{\widehat{a}^{(3)}}=-(m_{\psi}\sqrt{u^2}+a\cdot u)\,.
\end{equation}
The energy as a function of velocity reads
\begin{equation}
\label{eq:energy-velocity-a3}
E^{\widehat{a}^{(3)}}=\frac{m_{\psi}}{\sqrt{1-\mathbf{v}^2}}+a^{(3)}_0=m_{\psi}\left(1+\frac{1}{2}\mathbf{v}^2\right)+a^{(3)}_0+\dots\,,
\end{equation}
where the latter result is exact in Lorentz violation and valid at second order in the velocity. The momentum is then given by
\begin{equation}
p^i=\frac{m_{\psi}v^i}{\sqrt{1-\mathbf{v}^2}}-a_i\,.
\end{equation}
This is solved with respect to the velocity,
\begin{equation}
\label{eq:velocity-momentum-a3}
v^i=\frac{p^i+a_i}{\sqrt{(\mathbf{p}+\mathbf{a})^2+m_{\psi}^2}}=\frac{p^i+a_i}{m_{\psi}}-\frac{1}{2m_{\psi}^3}\left[\mathbf{p}^2a_i+2(\mathbf{a}\cdot \mathbf{p})p^i\right]+\dots\,,
\end{equation}
and it is inserted into \eqref{eq:energy-velocity-a3} to give the Hamilton function
\begin{equation}
\label{eq:energy-momentum-a3}
\mathcal{H}^{\widehat{a}^{(3)}}=m_{\psi}+\frac{\mathbf{p}^2}{2m_{\psi}}+a^{(3)}_0+\frac{\mathbf{a}\cdot\mathbf{p}}{m_{\psi}}+\dots\,.
\end{equation}
Note that a term comprising the scalar product $\mathbf{a}\cdot\mathbf{p}$ has emerged where there is no equivalent to such a
term in \eqref{eq:energy-velocity-a3}. This demonstrates that it is crucial to keep track of the contributions at first order
in Lorentz violation in the velocity-momentum correspondence of \eqref{eq:velocity-momentum-a3}.

\subsection{Operator \texorpdfstring{$\boldsymbol{\widehat{c}^{\,(4)}}$}{c4-hat}}

The Lagrange function associated to the dimensionless observer four-tensor coefficients $c^{(4)}_{\mu\nu}$ shall be considered
next where we define $\mathbf{c}\equiv (c^{(4)}_{01},c^{(4)}_{02},c^{(4)}_{03})$. The exact
result can be found in Eq.~(10) of \cite{Kostelecky:2010hs} and it involves both the symmetric and the antisymmetric part of
$c^{(4)}_{\mu\nu}$. However, the antisymmetric part contributes at second order in Lorentz violation only. Since in this article
all considerations are restricted to first order in Lorentz violation we start with the first-order expansion of the latter
Lagrange function, which is given by
\begin{equation}
L^{\widehat{c}^{\,(4)}}=-m_{\psi}\left(\sqrt{u^2}-\frac{c^{(4)}_{\mu\nu}u^{\mu}u^{\nu}}{\sqrt{u^2}}\right)+\dots\,.
\end{equation}
Here only the symmetric part of $c^{(4)}_{\mu\nu}$ contributes as expected. The particle energy as a function of the velocity
then reads as follows:
\begin{align}
\label{eq:energy-velocity-c4}
E^{\widehat{c}^{\,(4)}}&=\frac{m_{\psi}}{\sqrt{1-\mathbf{v}^2}}+\frac{m_{\psi}}{(1-\mathbf{v}^2)^{3/2}}\left[c^{(4)}_{ij}v^iv^j+2(\mathbf{c}\cdot\mathbf{v})\mathbf{v}^2-(1-2\mathbf{v}^2)c^{(4)}_{00}\right] \notag \\
&=m_{\psi}(1-c^{(4)}_{00})+\frac{1}{2}m_{\psi}\mathbf{v}^2(1+c^{(4)}_{00})+m_{\psi}c^{(4)}_{ij}v^iv^j+\dots\,.
\end{align}
Note that the latter does not involve the mixed components $c^{(4)}_{0i}$ with a timelike and a spacelike index. This is different
for the equations relating the spatial momentum to the velocity:
\begin{subequations}
\begin{align}
p^i&=m_{\psi}(1+c^{(4)}_{00})v^i+2m_{\psi}c^{(4)}_{0i}+2m_{\psi}c^{(4)}_{ij}v^j+\dots\,, \\[2ex]
v^i&=\frac{p^i}{m_{\psi}}(1-c^{(4)}_{00})-2c^{(4)}_{0i}-\frac{2}{m_{\psi}}c^{(4)}_{ij}p^j+\dots\,.
\end{align}
\end{subequations}
Therefore, replacing the velocity by the momentum in \eqref{eq:energy-velocity-c4} introduces coefficients $c^{(4)}_{0i}$ into
the Hamilton function:
\begin{equation}
\label{eq:energy-momentum-c4}
\mathcal{H}^{\widehat{c}^{\,(4)}}=(1-c^{(4)}_{00})\left(m_{\psi}+\frac{\mathbf{p}^2}{2m_{\psi}}\right)-2(\mathbf{c}\cdot\mathbf{p})-\frac{1}{m_{\psi}}c^{(4)}_{ij}p^ip^j+\dots\,.
\end{equation}

\subsection{Operator \texorpdfstring{$\boldsymbol{\widehat{e}^{\,(4)}}$}{e4-hat}}

The next case to be studied is the observer four-vector $e^{(4)}_{\mu}=(e^{(4)}_0,\mathbf{e})_{\mu}$ with $\mathbf{e}\equiv (e^{(4)}_1,e^{(4)}_2,e^{(4)}_3)$
including dimensionless coefficients. The Lagrangian follows from Eq.~(8) in \cite{Kostelecky:2010hs} by setting the $a$ and $f$ coefficients to zero:
\begin{equation}
L^{\widehat{e}^{\,(4)}}=m_{\psi}(-\sqrt{u^2}+e\cdot u)\,,
\end{equation}
where its structure is very similar to the Lagrangian of $a^{(3)}_{\mu}$ given in \eqref{eq:lagrangian-a3}. The particle energy
as a function of velocity reads
\begin{equation}
\label{energy-velocity-e4}
E^{\widehat{e}^{\,(4)}}=m_{\psi}\left(\frac{1}{\sqrt{1-\mathbf{v}^2}}-e^{(4)}_0\right)=m_{\psi}\left(1+\frac{1}{2}\mathbf{v}^2\right)-m_{\psi}e^{(4)}_0+\dots\,.
\end{equation}
It does not involve the spatial components of $e^{(4)}_{\mu}$, which is a behavior similar to
\eqref{eq:energy-velocity-a3} that does not comprise the spatial components of $a^{(3)}_{\mu}$ either. The correspondences between velocity and momentum read
\begin{subequations}
\begin{align}
p^i&=m_{\psi}\left(\frac{v^i}{\sqrt{1-\mathbf{v}^2}}+e^{(4)}_i\right)=m_{\psi}(v^i+e^{(4)}_i)+\dots\,, \\[2ex]
v^i&=\frac{p^i-m_{\psi}e^{(4)}_i}{\sqrt{\mathbf{p}^2-2m_{\psi}\mathbf{e}\cdot\mathbf{p}+m_{\psi}^2(1+\mathbf{e}^2)}}=\frac{1}{m_{\psi}}(p^i-m_{\psi}e^{(4)}_i)\left(1+\frac{1}{m_{\psi}}\mathbf{e}\cdot\mathbf{p}-\frac{\mathbf{p}^2}{2m_{\psi}^2}\right)+\dots \notag \\
&=\frac{p^i}{m_{\psi}}-e_i^{(4)}+\frac{p^i}{m_{\psi}^2}(\mathbf{e}\cdot\mathbf{p})+\frac{\mathbf{p}^2}{2m_{\psi}^2}e_i^{(4)}+\dots\,.
\end{align}
\end{subequations}
Replacing the velocity by the momentum in \eqref{energy-velocity-e4} introduces the spatial components of $e^{(4)}_{\mu}$ into the energy,
which works in analogy to the case of the $a$ coefficients:
\begin{equation}
\label{eq:energy-momentum-e4}
\mathcal{H}^{\widehat{e}^{\,(4)}}=m_{\psi}(1-e^{(4)}_0)+\frac{\mathbf{p}^2}{2m_{\psi}}-\mathbf{e}\cdot\mathbf{p}+\dots
\end{equation}
These results demonstrate that the families of the minimal $a$ and $e$ coefficients behave in a very similar manner.
This is not surprising since the $a$ and $e$ coefficients are part of the same effective coefficient, cf.~the first
of Eq.~(27) in \cite{Kostelecky:2013rta}.

\subsection{Operator \texorpdfstring{$\boldsymbol{\widehat{b}^{(3)}}$}{b3-hat}}

The three frameworks previously considered do not break degeneracy with respect to the particle spin, i.e., there is only a
single classical Lagrangian corresponding to the particle solutions in quantum field theory. For the following cases this
degeneracy is broken, starting with the $b^{(3)}_{\mu}$ coefficients of mass dimension 1. The Lagrangian is obtained from Eq.~(12)
in \cite{Kostelecky:2010hs} by setting the $a$ coefficients to zero:
\begin{equation}
\label{eq:lagrangian-b3}
L^{\widehat{b}^{(3)}\pm}=-m_{\psi}\sqrt{u^2}\mp\sqrt{(b\cdot u)^2-b^2u^2}\,.
\end{equation}
The two signs break the spin degeneracy as mentioned. The upper sign always corresponds to the configuration of ``spin-up''
and the lower to ``spin-down.'' Although the concept of spin does not exist classically this correspondence can be inferred
from the quantum theory. We will come back to this point later. Note that the Lorentz-violating contribution has a very different
structure compared to the cases of the $a$, $e$, and $c$ coefficients; it is called bipartite \cite{Kostelecky:2012ac}.
The Lagrangian was shown to be related to a Finsler
space that is neither Riemannian nor of Randers-type \cite{Kostelecky:2010hs}. One peculiarity is that it is not straightforward
to expand \eqref{eq:lagrangian-b3} with respect to the controlling coefficients or the velocity. Therefore, we consider two
observer frames: the first with $b_{\mu}^{(3)}$ being purely spacelike and the second with $b_{\mu}^{(3)}$ purely timelike.

\subsubsection{Spacelike part}
\label{sec:b-space-spacelike}

The first case is based on a purely spacelike observer four-vector $b_{\mu}^{(3)}$ that is expressed as $b_{\mu}^{(3)}=(0,\mathbf{b})_{\mu}$
with the spatial part $\mathbf{b}\equiv (b_1^{(3)},b_2^{(3)},b_3^{(3)})$ where the latter will occur in all results.
For such a choice the Lagrangian of \eqref{eq:lagrangian-b3} reads as follows:
\begin{equation}
L^{\widehat{b}^{(3)}\pm}_1=-m_{\psi}\sqrt{u^2}\mp\sqrt{\mathbf{b}^2u^2+(\mathbf{b}\cdot\mathbf{u})^2}\,.
\end{equation}
The energy can then be obtained just as before by differentiating the Lagrangian with respect to $u^0$ or by a Legendre transformation.
The result shall be expanded at first order in Lorentz violation. This is a bit more involved compared to the previous
cases since the Lorentz-violating contribution behaves asymptotically like a square root function that does not have
a Taylor expansion for a vanishing argument. This can be remedied by introducing the angle $\theta$ between the velocity
and the three-vector $\mathbf{b}$ composed of the controlling coefficients. After doing so, the magnitude of $\mathbf{b}$
can be extracted from the expression which allows for a subsequent expansion with respect to the velocity:
\begin{align}
\label{eq:energy-velocity-b3}
E^{\widehat{b}^{(3)}\pm}_1&=\frac{m_{\psi}}{\sqrt{1-\mathbf{v}^2}}\pm\frac{\mathbf{b}^2}{\sqrt{(1-\mathbf{v}^2)\mathbf{b}^2+(\mathbf{b}\cdot\mathbf{v})^2}}=\frac{m_{\psi}}{\sqrt{1-\mathbf{v}^2}}\pm\frac{|\mathbf{b}|}{\sqrt{1-\mathbf{v}^2\sin^2\theta}} \notag \\
&=m_{\psi}\left(1+\frac{1}{2}\mathbf{v}^2\right)\pm|\mathbf{b}|\left(1+\frac{1}{2}\mathbf{v}^2\sin^2\theta\right)+\dots\,.
\end{align}
We apply the same method to perform expansions of the momentum given as a function of the velocity:
\begin{equation}
\label{eq:momentum-velocity-b3}
p^i=\frac{m_{\psi}v^i}{\sqrt{1-\mathbf{v}^2}}\pm\frac{\mathbf{b}^2v^i-b_i(\mathbf{b}\cdot\mathbf{v})}{\sqrt{(1-\mathbf{v}^2)\mathbf{b}^2+(\mathbf{b}\cdot\mathbf{v})^2}}=(m_{\psi}\pm|\mathbf{b}|)v^i\mp b_i|\mathbf{v}|\cos\theta+\dots\,.
\end{equation}
The latter is then solved for the velocity components to give
\begin{equation}
\label{eq:velocity-momentum-b3}
v^i=\frac{p^i}{m_{\psi}}\left(1\mp\frac{|\mathbf{b}|}{m_{\psi}}\right)\pm b_i\frac{|\mathbf{p}|}{m_{\psi}^2}\cos\theta+\dots\,.
\end{equation}
At this point there is a subtle issue that occurs for the first time in the course of our studies. When solving
\eqref{eq:momentum-velocity-b3} for the velocity we encounter an absolute value of the trigonometric function in
\eqref{eq:velocity-momentum-b3}. Eliminating these absolute-value bars would lead to four different sign choices
dependent on both the angle $\theta$ and the upper or lower sign coming from the original Lagrangian. In the end this would
result in four different Hamilton functions, which does not match the number of degrees of freedom in the original
Lagrangian. Therefore, we have to set up a proposal telling us how to choose the signs appropriately to obtain two distinct
Hamilton functions corresponding to the two Lagrangians given initially.

\begin{figure}
\centering
\includegraphics{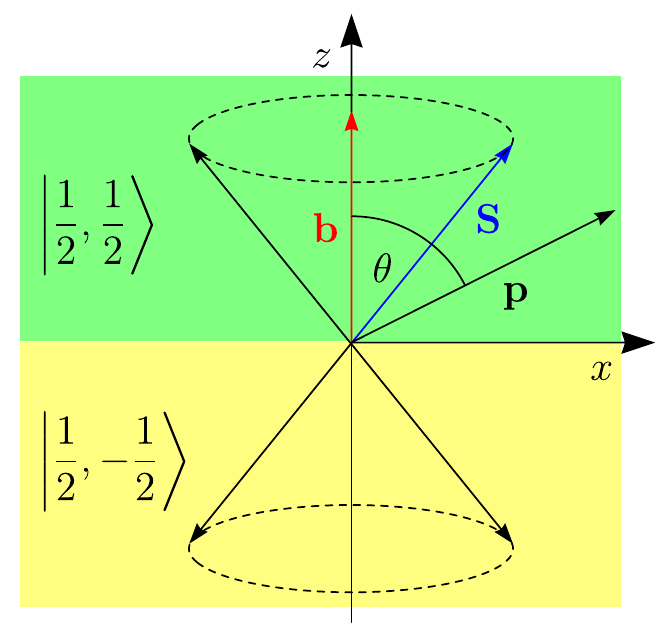}
\caption{Coordinate frame including the preferred direction $\mathbf{b}$, the
particle velocity $\mathbf{v}$, and the angle $\theta$ between the preferred direction and the velocity. The spin vector
$\mathbf{S}$ is shown as well with the quantization axis chosen to be the $z$ axis. By doing so, the spin-up state $|1/2,1/2\rangle$
is situated in the upper half-plane and the spin-down state $|1/2,-1/2\rangle$ is realized in the lower half-plane. Hence,
angles $\theta\in [0,\pi/2]$ are connected to the spin pointing up and angles $\theta\in (\pi/2,\pi]$ to the spin pointing
down.}
\label{fig:sign-choice-momentum-correspondence}
\end{figure}

On the base of observer Lorentz invariance, a coordinate system is defined such that its $z$ axis points along the preferred
direction $\mathbf{b}$. The spin quantization axis can be chosen freely and for convenience it is arranged to point along
the $z$ axis as well, cf.~\figref{fig:sign-choice-momentum-correspondence}. One the one hand, the spin-up state
$|1/2,1/2\rangle$ can then be understood to be realized in the upper half-plane. That corresponds to $\theta\in [0,\pi/2]$
for the angle $\theta$ between the particle velocity $\mathbf{v}$ and the preferred axis. On the other hand, the
spin-down state $|1/2,-1/2\rangle$ is realized in the lower half-plane where $\theta\in (\pi/2,\pi]$. Now assume that
the particle is in a spin-up state. For $\theta\in [0,\pi/2)$ the absolute value bars around $\cos\theta$ do not have any
effect, which is why they will be dropped, choosing the upper of the two signs in \eqref{eq:velocity-momentum-b3}. For
$\theta\in (\pi/2,\pi]$ the absolute value bars act like a minus sign before the term affected. Therefore, upon dropping
them, the lower sign is picked. So each sign is not valid for all momentum configurations possible but just
for a restricted range of angles. For the particle being in a spin-down state, the same procedure can be applied with
both signs switched. Then for $\theta\in [0,\pi/2)$ the minus sign must be chosen and for $\theta\in (\pi/2,\pi]$
the plus sign.

Finally, in \eqref{eq:energy-velocity-b3} the velocity is replaced by the momentum leading to the Hamilton function:
\begin{align}
\label{eq:energy-momentum-b3}
\mathcal{H}^{\widehat{b}^{(3)}\pm}_1&=m_{\psi}+\frac{\mathbf{p}^2}{2m_{\psi}}\pm |\mathbf{b}|\left(1-\frac{\mathbf{p}^2}{2m_{\psi}^2}\sin^2\theta\right)+\dots \notag \\
&=m_{\psi}+\frac{\mathbf{p}^2}{2m_{\psi}}\pm\left(|\mathbf{b}|-\frac{1}{2m_{\psi}^2}\frac{(\mathbf{b}\times\mathbf{p})^2}{|\mathbf{b}|}\right)+\dots\,.
\end{align}
In the last step we introduced the cross product between the spatial momentum and the spacelike vector $\mathbf{b}$
of the controlling coefficients. Note that in general the velocity vector does not point along the direction of the
momentum vector, cf.~Eqs.~(\ref{eq:momentum-velocity-b3}), (\ref{eq:velocity-momentum-b3}). Hence, the angle between
$\mathbf{p}$ and $\mathbf{b}$ deviates from the angle $\theta$ between $\mathbf{v}$ and $\mathbf{b}$. However, the
deviation is of first order in Lorentz violation (see, e.g., Eqs.~(13), (16) in \cite{Kostelecky:2010hs} for the $b$
and $H$ coefficients, respectively), which leads to second-order corrections in \eqref{eq:energy-momentum-b3} that
are discarded anyhow. With the spin quantization axis pointing along the $z$ axis and the particle being in a spin-up
state, the upper sign of the Hamilton function must be picked for $\theta\in [0,\pi/2]$ and the lower sign for
$\theta\in (\pi/2,\pi]$. When the particle is in a spin-down state, both signs have to be switched.

\subsubsection{Timelike part}
\label{sec:b-space-timelike}

To study the second case of the $b$ coefficients an observer frame is chosen where $b^{(3)}_{\mu}$ is purely timelike.
Such a choice involves a single controlling coefficient: $b^{(3)}_{\mu}=(b_0^{(3)},\mathbf{0})_{\mu}$.
The Lagrangian is isotropic and takes a simple form:
\begin{equation}
L^{\widehat{b}^{(3)}\pm}_2=-m_{\psi}\sqrt{u^2}\mp |b^{(3)}_0||\mathbf{u}|\,.
\end{equation}
Since the Lorentz-violating contribution does not comprise $u^0$ the energy corresponds to the
standard expression when expressed in terms of the velocity:
\begin{equation}
\label{eq:energy-velocity-b30}
E^{\widehat{b}^{(3)}\pm}_2=\frac{m_{\psi}}{\sqrt{1-\mathbf{v}^2}}=m_{\psi}\left(1+\frac{1}{2}\mathbf{v}^2\right)+\dots
\end{equation}
However, this is not the case for the momentum since the latter is obtained as the first derivative of the Lagrangian
with respect to $\mathbf{u}$. In proper-time parameterization we obtain
\begin{equation}
p^i=v^i\left(\frac{m_{\psi}}{\sqrt{1-\mathbf{v}^2}}\mp\frac{|b^{(3)}_0|}{|\mathbf{v}|}\right)=m_{\psi}\left(1+\frac{1}{2}\mathbf{v}^2\right)v^i\mp |b^{(3)}_0|\frac{v^i}{|\mathbf{v}|}+\dots\,.
\end{equation}
Solving this relation for the velocity leads to
\begin{equation}
v^i=\frac{p^i}{m_{\psi}}\left(1\pm\frac{|b^{(3)}_0|}{|\mathbf{p}|}\right)+\dots\,.
\end{equation}
Note the singularities in $|\mathbf{v}|$ and $|\mathbf{p}|$ in the latter two expressions. Furthermore, here it must
be distinguished between the two signs: the upper sign has to be chosen for $p^1\geq 0$ and the lower for $p^1<0$.
This procedure differs from the prescription that we introduced in the last section. It is challenging to illustrate
it physically by taking into account the particle spin just as we did in \secref{sec:b-space-spacelike}. It seems
that a similar procedure always has to be carried out when there are singularities in $|\mathbf{v}|$, $|\mathbf{p}|$ in the
momentum-velocity correspondences, cf. the forthcoming  Secs.~\ref{sec:H-space-case-2}, \ref{sec:d-space-case-3}.
Now, replacing $\mathbf{v}$ by $\mathbf{p}$ in \eqref{eq:energy-velocity-b30} introduces the single controlling
coefficient into the Hamilton function:
\begin{equation}
\mathcal{H}^{\widehat{b}^{(3)}\pm}_2=m_{\psi}+\frac{\mathbf{p}^2}{2m_{\psi}}\pm |b^{(3)}_0|\frac{|\mathbf{p}|}{m_{\psi}}+\dots\,.
\end{equation}
The resulting expression is isotropic, as expected, and does not have any singularities.

\subsection{Operator \texorpdfstring{$\boldsymbol{\widehat{H}^{(3)}}$}{H3-hat}}

The next cases to be studied involve the observer two-tensor coefficients $H^{(3)}_{\mu\nu}$ that are of mass dimension
1. Several Lagrangians valid for particular subsets of coefficients were obtained in the literature. In this context
the following observer Lorentz scalars are helpful:
\begin{subequations}
\begin{align}
\alpha&\equiv (u\cdot H^{(3)})_{\mu}(H^{(3)}\cdot u)^{\mu}\,, \\[2ex]
X&\equiv \frac{1}{4}H^{(3)}_{\mu\nu}H^{(3)\mu\nu}\,,\quad Y\equiv \frac{1}{4}H^{(3)}_{\mu\nu}\widetilde{H}^{(3)\mu\nu}\,,\quad \widetilde{H}^{(3)\mu\nu}\equiv \frac{1}{2}\varepsilon^{\mu\nu\varrho\sigma}H^{(3)}_{\varrho\sigma}\,,
\end{align}
\end{subequations}
where $\varepsilon^{\mu\nu\varrho\sigma}$ with $\varepsilon^{0123}=1$ is the totally antisymmetric Levi-Civita symbol
in four spacetime dimensions. The matrix $H^{(3)}_{\mu\nu}$ can be taken as antisymmetric, cf.~\cite{Kostelecky:2000mm}.

\subsubsection{Spacelike case with \texorpdfstring{$X\neq 0$}{x} and \texorpdfstring{$Y=0$}{y}}
\label{sec:H-space-case-1}

The first Lagrangian considered is valid for $X\neq 0$ and $Y=0$. It is given by Eq.~(15) in \cite{Kostelecky:2010hs}:
\begin{equation}
\label{eq:lagrangian-H3-first}
L^{\widehat{H}^{(3)}\pm}_1=-m_{\psi}\sqrt{u^2}\pm\sqrt{\alpha+2Xu^2}\,.
\end{equation}
Spin degeneracy is again broken just as for the $b$ coefficients. An important choice that fulfills $X\neq 0$
and $Y=0$ is an antisymmetric $H^{(3)}_{\mu\nu}$ comprised solely of controlling coefficients with spacelike indices:
\begin{subequations}
\label{eq:preferred-direction-H3-first}
\begin{align}
H^{(3)}_{\mu\nu}&=\begin{pmatrix}
0 & 0 & 0 & 0 \\
0 & 0 & h^{(3)}_{12} & h^{(3)}_{13} \\
0 & -h^{(3)}_{12} & 0 & h^{(3)}_{23} \\
0 & -h^{(3)}_{13} & -h^{(3)}_{23} & 0 \\
\end{pmatrix}_{\mu\nu}\,,\quad \mathbf{h}\equiv\begin{pmatrix}
h^{(3)}_{23} \\
-h^{(3)}_{13} \\
h^{(3)}_{12} \\
\end{pmatrix}\,, \\[2ex]
\alpha&=\mathbf{h}^2\mathbf{u}^2-(\mathbf{h}\cdot\mathbf{u})^2\,,\quad X=\frac{1}{2}\mathbf{h}^2\,,\quad Y=0\,.
\end{align}
\end{subequations}
The energy bears some similarities to \eqref{eq:energy-velocity-b3}. Expansions with respect to the controlling
coefficients and the velocity are computed as before. We introduce the angle $\theta$ between the velocity vector
and the vector $\mathbf{h}$ comprising the controlling coefficients. Then $|\mathbf{h}|$ is extracted from the square root
and the resulting expression is expanded with respect to the velocity:
\begin{align}
\label{eq:energy-H3-first}
E^{\widehat{H}^{(3)}\pm}_1&=\frac{m_{\psi}}{\sqrt{1-\mathbf{v}^2}}\mp\frac{\mathbf{h}^2}{\sqrt{\mathbf{h}^2-(\mathbf{h}\cdot\mathbf{v})^2}}=\frac{m_{\psi}}{\sqrt{1-\mathbf{v}^2}}\mp\frac{|\mathbf{h}|}{\sqrt{1-\mathbf{v}^2\cos^2\theta}} \notag \\
&=m_{\psi}\left(1+\frac{1}{2}\mathbf{v}^2\right)\mp |\mathbf{h}|\left(1+\frac{1}{2}\mathbf{v}^2\cos^2\theta\right)+\dots
\end{align}
Similarly, this procedure is applied to obtain the momentum:
\begin{align}
\label{eq:momentum-velocity-correspondence-H3-first}
p^i&=\frac{m_{\psi}v^i}{\sqrt{1-\mathbf{v}^2}}\mp\frac{h_i(\mathbf{h}\cdot\mathbf{v})}{\sqrt{\mathbf{h}^2-(\mathbf{h}\cdot\mathbf{v})^2}}=\frac{m_{\psi}v^i}{\sqrt{1-\mathbf{v}^2}}\mp h_i\frac{|\mathbf{v}|\cos\theta}{\sqrt{1-\mathbf{v}^2\cos^2\theta}} \notag \\
&=m_{\psi}v^i\mp h_i|\mathbf{v}|\cos\theta+\dots\,,
\end{align}
which is then solved for the velocity:
\begin{equation}
\label{eq:velocity-momentum-correspondence-H3-first}
v^i=\frac{p^i}{m_{\psi}}\pm h_i\frac{|\mathbf{p}|}{m_{\psi}^2}\cos\theta+\dots\,.
\end{equation}
Here the same issue appears that we encountered for the spacelike case of the $b$ coefficients in \secref{sec:b-space-spacelike},
i.e., the sign choice depends on the angle $\theta$. When the spin quantization axis points along the $z$ direction and the
particle is in a spin-up state, we choose the upper sign for $\theta\in [0,\pi/2]$ and the lower for $\theta\in (\pi/2,\pi]$.
The signs must be picked vice versa for the particle being in a spin-down state. Plugging the velocity-momentum correspondence of
\eqref{eq:velocity-momentum-correspondence-H3-first} into \eqref{eq:energy-H3-first}, the final result is the Hamilton function:
\begin{align}
\label{eq:energy-momentum-H3}
\mathcal{H}^{\widehat{H}^{(3)}\pm}_1&=m_{\psi}+\frac{\mathbf{p}^2}{2m_{\psi}}\mp|\mathbf{h}|\left(1-\frac{\mathbf{p}^2}{2m_{\psi}^2}\cos^2\theta\right)+\dots \notag \\
&=m_{\psi}+\frac{\mathbf{p}^2}{2m_{\psi}}\mp\left(|\mathbf{h}|-\frac{1}{2m_{\psi}^2}\frac{(\mathbf{h}\cdot\mathbf{p})^2}{|\mathbf{h}|}\right)+\dots\,.
\end{align}
In the last step we expressed the result by the scalar product of $\mathbf{h}$ and $\mathbf{p}$ in analogy to how we dealt with
\eqref{eq:energy-momentum-b3} by introducing the cross product. Note that the angle $\theta$ does not correspond to
the angle between $\mathbf{h}$ and $\mathbf{p}$. However, deviations are of first order in Lorentz violation, which
is why those produce higher-order terms in the final result of \eqref{eq:energy-momentum-H3}. For spin pointing up, the upper
sign of the Hamilton function holds for $\theta\in [0,\pi/2]$ and the lower sign for $\theta\in (\pi/2,\pi]$,
cf.~\secref{sec:b-space-spacelike}. For spin pointing down, the opposite is the case.

\subsubsection{Timelike case with \texorpdfstring{$X\neq 0$}{x} and \texorpdfstring{$Y=0$}{y}}
\label{sec:H-space-case-2}

Another case with $Y=0$ is constructed from an antisymmetric $H^{(3)}_{\mu\nu}$ with nonzero coefficients having one
timelike index. This choice reads as follows:
\begin{subequations}
\begin{align}
H^{(3)}_{\mu\nu}&=\begin{pmatrix}
0 & h^{(3)}_{01} & h^{(3)}_{02} & h^{(3)}_{03} \\
-h^{(3)}_{01} & 0 & 0 & 0 \\
-h^{(3)}_{02} & 0 & 0 & 0 \\
-h^{(3)}_{03} & 0 & 0 & 0 \\
\end{pmatrix}_{\mu\nu}\,,\quad \mathbf{h}\equiv\begin{pmatrix}
h^{(3)}_{01} \\
h^{(3)}_{02} \\
h^{(3)}_{03} \\
\end{pmatrix}\,, \\[2ex]
\alpha&=\mathbf{h}^2(u^0)^2-(\mathbf{h}\cdot\mathbf{u})^2\,,\quad X=-\frac{1}{2}\mathbf{h}^2\,,\quad Y=0\,.
\end{align}
\end{subequations}
That sector is also based on the Lagrangian in \eqref{eq:lagrangian-H3-first}. The energy is obtained from its first derivative
with respect to $u^0$:
\begin{equation}
E^{\widehat{H}^{(3)}}_2=\frac{m_{\psi}}{\sqrt{1-\mathbf{v}^2}}=m_{\psi}\left(1+\frac{1}{2}\mathbf{v}^2\right)+\dots\,,
\end{equation}
and it does not involve any Lorentz-violating terms at first order. This is different for the momentum and the velocity
that are given by:
\begin{subequations}
\begin{align}
p^i&=\frac{m_{\psi}v^i}{\sqrt{1-\mathbf{v}^2}}\pm\frac{\mathbf{h}^2v^i-h_i(\mathbf{h}\cdot\mathbf{v})}{\sqrt{\mathbf{h}^2\mathbf{v}^2-(\mathbf{h}\cdot\mathbf{v})^2}}=m_{\psi}v^i\pm|\mathbf{h}|\frac{v^i}{|\mathbf{v}|}\csc\theta\mp h_i\cot\theta+\dots\,, \\[2ex]
v^i&=\frac{p^i}{m_{\psi}}\left(1\mp \frac{|\mathbf{h}|}{|\mathbf{p}|}\csc\theta\right)\pm \frac{h_i}{m_{\psi}}\cot\theta+\dots\,.
\end{align}
\end{subequations}
Both expressions are singular in $|\mathbf{v}|$ and $|\mathbf{p}|$, respectively. Besides, the upper sign holds for $p^1\geq 0$
and the lower for $p^1<0$. Recall that the timelike case of the $b$ coefficients behaved similarly, cf.~\secref{sec:b-space-timelike}.
The singularities do not occur in the Hamilton function, though:
\begin{equation}
\mathcal{H}^{\widehat{H}^{(3)}}_2=m_{\psi}+\frac{\mathbf{p}^2}{2m_{\psi}}\mp \frac{1}{m_{\psi}}|\mathbf{h}||\mathbf{p}|\sin\theta+\dots=m_{\psi}+\frac{\mathbf{p}^2}{2m_{\psi}}\mp \frac{1}{m_{\psi}}|\mathbf{h}\times\mathbf{p}|+\dots
\end{equation}
Just as before, we introduce the cross product between $\mathbf{h}$ and the three-momentum neglecting higher-order
terms in Lorentz violation.

\subsubsection{Case with \texorpdfstring{$X=0$}{x} and \texorpdfstring{$Y\neq 0$}{y}}
\label{sec:H-space-case-3}

The case of $H^{(3)}_{\mu\nu}$ with $X=0$ and $Y\neq 0$ was considered in \cite{Kostelecky:2010hs} as well. This particular
framework is much more complicated than the previous one, which becomes manifest in the polynomial of their Eq.~(18) whose
(perturbative) zeros with respect to $L$ correspond to the Lagrange functions searched for. In \cite{Kostelecky:2010hs} they
are not stated explicitly due to their complicated structure. Here we consider the following configuration of nonvanishing
controlling coefficients:
\begin{subequations}
\begin{align}
h^{(3)}_{01}&\equiv h\equiv -h^{(3)}_{10}\,,\quad h^{(3)}_{23}\equiv h\equiv -h^{(3)}_{32}\,, \\[2ex]
\label{eq:H3-case-2-preferred-directions}
H^{(3)}_{\mu\nu}&=\xi_{\mu}\zeta_{\nu}-\zeta_{\mu}\xi_{\nu}+\varepsilon_{\mu\nu}^{\phantom{\mu\nu}\varrho\sigma}\xi_{\varrho}\zeta_{\sigma}\,,\quad (\xi_{\mu})=\begin{pmatrix}
1 \\
\mathbf{0} \\
\end{pmatrix}\,,\quad (\zeta_{\mu})=\begin{pmatrix}
0 \\
\mathbf{h} \\
\end{pmatrix}\,,\quad \mathbf{h}=\begin{pmatrix}
h \\
0 \\
0 \\
\end{pmatrix}\,, \\[2ex]
\alpha&=h^2[(u^0)^2-(u^1)^2+(u^2)^2+(u^3)^2]=h^2\left[2(u\cdot\xi)^2-u^2\right]-2(\zeta\cdot u)^2\,, \\[2ex]
X&=0\,,\quad Y=\mathbf{h}^2=h^2\,.
\end{align}
\end{subequations}
Evidently, $H^{(3)}_{\mu\nu}$ can be expressed by a timelike preferred direction $\xi_{\mu}$ and a spacelike one
$\zeta_{\mu}$, i.e., these directions are physical and especially the spacelike one will appear in the results below.
For this choice the Lagrange functions are obtained from Eq.~(18) of \cite{Kostelecky:2010hs} with computer algebra where
a subsequent expansion of the result at first order in Lorentz violation leads to
\begin{equation}
\label{eq:lagrangian-H3-second}
L^{\widehat{H}^{(3)}\pm}_3=-m_{\psi}\sqrt{u^2}\pm \sqrt{\alpha}+\mathcal{O}(h^2)\,.
\end{equation}
The Lorentz-violating contribution can be expressed by the observer Lorentz scalar $\alpha$. This means that the form of
the result stays unchanged in an arbitrary observer frame that can be transformed to by an observer Lorentz transformation.
At first order in Lorentz violation, the Lagrangian corresponds to \eqref{eq:lagrangian-H3-first} for $X=0$.
The energy is then given by
\begin{align}
\label{eq:energy-velocity-H3-second}
E^{\widehat{H}^{(3)}\pm}_3&=\frac{m_{\psi}}{\sqrt{1-\mathbf{v}^2}}\mp\frac{|\mathbf{h}|}{\sqrt{1-(v^1)^2+(v^2)^2+(v^3)^2}} \notag \\
&=m_{\psi}\left(1+\frac{1}{2}\mathbf{v}^2\right)\mp \left[|\mathbf{h}|-\frac{1}{2}\left(\frac{(\mathbf{h}\times\mathbf{v})^2}{|\mathbf{h}|}-\frac{(\mathbf{h}\cdot\mathbf{v})^2}{|\mathbf{h}|}\right)\right]\,,
\end{align}
which can be written in terms of the spatial part $\mathbf{h}$ of the second preferred direction,
cf.~\eqref{eq:H3-case-2-preferred-directions}. The momentum-velocity correspondence reads
\begin{subequations}
\begin{align}
p^1&=(m_{\psi}\mp |\mathbf{h}|)v^1+\dots\,,\quad p^i=(m_{\psi}\pm |\mathbf{h}|)v^i+\dots\,, \\[2ex]
v^1&=\frac{p^1}{m_{\psi}}\left(1\pm\frac{|\mathbf{h}|}{m_{\psi}}\right)+\dots\,,\quad v^i=\frac{p^i}{m_{\psi}}\left(1\mp\frac{|\mathbf{h}|}{m_{\psi}}\right)+\dots\,,
\end{align}
\end{subequations}
where $i\in\{2,3\}$. From this we obtain the Hamilton function:
\begin{align}
\label{eq:energy-momentum-H3-second}
\mathcal{H}^{\widehat{H}^{(3)}}_3&=m_{\psi}+\frac{\mathbf{p}^2}{2m_{\psi}}\mp |\mathbf{h}|\left(1+\frac{-(p^1)^2+(p^2)^2+(p^3)^2}{2m_{\psi}^2}\right)+\dots \notag \\
&=m_{\psi}+\frac{\mathbf{p}^2}{2m_{\psi}}\mp \left[|\mathbf{h}|+\frac{1}{2m_{\psi}^2}\left(\frac{(\mathbf{h}\times\mathbf{p})^2}{|\mathbf{h}|}-\frac{(\mathbf{h}\cdot\mathbf{p})^2}{|\mathbf{h}|}\right)\right]+\dots\,.
\end{align}
Note the similar structure of the result in comparison to \eqref{eq:energy-velocity-H3-second}.
As mentioned above, the Lagrangian of \eqref{eq:lagrangian-H3-second} can be transformed to another observer frame.
The special $\mathbf{h}$ given in \eqref{eq:H3-case-2-preferred-directions} points along the first spatial direction of the
coordinate frame. The general case results from an observer rotation such that $\mathbf{h}$ points along an arbitrary
direction.

\subsection{Operator \texorpdfstring{$\boldsymbol{\widehat{d}^{\,(4)}}$}{d4-hat}}

The $d$ coefficients are dimensionless and comprised by an observer two-tensor $d^{(4)}_{\mu\nu}$ that can be taken
as traceless, cf.~\cite{Kostelecky:2000mm}. The corresponding Lagrangian is challenging to be obtained since the
dispersion relation is quartic in $p_0$. Nevertheless, two cases were considered in \cite{Colladay:2012rv}. For convenience,
we define a couple of helpful observer Lorentz scalars as follows:
\begin{subequations}
\begin{align}
D&\equiv (u\cdot d^{(4)})_{\mu}(d^{(4)}\cdot u)^{\mu}\,, \\[2ex]
X&\equiv \frac{1}{4}d^{(4)}_{\mu\nu}d^{(4)\mu\nu}\,,\quad Y\equiv \frac{1}{4}d_{\mu\nu}^{(4)}\widetilde{d}^{(4)\mu\nu}\,,\quad \widetilde{d}^{(4)\mu\nu}\equiv\frac{1}{2}\varepsilon^{\mu\nu\varrho\sigma}d^{(4)}_{\varrho\sigma}\,.
\end{align}
\end{subequations}

\subsubsection{Nonsymmetric operator with nonvanishing timelike components}
\label{sec:d-space-case-1}

The first case involves the nonvanishing coefficients $d^{(4)}_{0i}$ only, i.e., the whole tensor is not symmetric.
We introduce the three-vector $\mathbf{d}$ including these coefficients, which turns out to be a useful quantity:
\begin{equation}
d^{(4)}_{\mu\nu}=\begin{pmatrix}
0 & d^{(4)}_{01} & d^{(4)}_{02} & d^{(4)}_{03} \\
0 & 0 & 0 & 0 \\
0 & 0 & 0 & 0 \\
0 & 0 & 0 & 0 \\
\end{pmatrix}_{\mu\nu}\,,\quad \mathbf{d}\equiv\begin{pmatrix}
d^{(4)}_{01} \\
d^{(4)}_{02} \\
d^{(4)}_{03} \\
\end{pmatrix}\,.
\end{equation}
The corresponding Lagrangian is given by Eqs.~(26), (27) in \cite{Colladay:2012rv}. Its form is rather complicated; it
comprises both a cross product and a scalar product of $\mathbf{d}$ and $\mathbf{v}$. To consider the expression
at leading order in Lorentz violation and at second order in the velocity we follow the procedure already used
for the $b$ and $H$ coefficients. Let $\theta$ be the angle between $\mathbf{d}$ and the velocity.
The Lagrangian then reads as follows:
\begin{subequations}
\label{eq:lagrangian-d4-first}
\begin{align}
L^{\widehat{d}^{\,(4)}\pm}_1&=-m_{\psi}\sqrt{1-\frac{\mathbf{v}^2}{C_{d\pm}}}\,, \\[2ex]
C_{d\pm}&=\left(\sqrt{1-(\widehat{\mathbf{v}}\times\mathbf{d})^2}\pm|\widehat{\mathbf{v}}\cdot\mathbf{d}|\right)^2=\left(\sqrt{1-\mathbf{d}^2\sin^2\theta}\pm |\mathbf{d}||\cos\theta|\right)^2\,,
\end{align}
\end{subequations}
with the velocity unit vector $\widehat{\mathbf{v}}\equiv \mathbf{v}/|\mathbf{v}|$. The Lagrangian was constructed
directly in proper-time parameterization, which is why it does not depend on $u^0$. Therefore, the energy cannot be
computed via \eqref{eq:energy-from-derivative-u0} but we have to perform a Legendre transformation according to
\eqref{eq:legendre-transform}. The result at first order in Lorentz violation is given by:
\begin{equation}
E^{\widehat{d}^{\,(4)}\pm}_1=\frac{m_{\psi}}{\sqrt{1-\mathbf{v}^2/C_{d\pm}}}=m_{\psi}\left(1+\frac{1}{2}\mathbf{v}^2\right)\mp m_{\psi}|\mathbf{d}\cdot\mathbf{v}||\mathbf{v}|+\dots\,.
\end{equation}
The momentum reads
\begin{align}
p^i&=\frac{m_{\psi}}{C_{d\pm}\sqrt{C_{d\pm}-\mathbf{v}^2}}\left[2\sqrt{C_{d\pm}}v^i\mp|\mathbf{v}|\left(|d_i|\pm\frac{(1-\mathbf{d}^2)v^i+d_i(\mathbf{d}\cdot\mathbf{v})}{\sqrt{(1-\mathbf{d}^2)\mathbf{v}^2+(\mathbf{d}\cdot\mathbf{v})^2}}\right)\right] \notag \\
&=m_{\psi}\left[(v^i\mp|\mathbf{d}||v^i\cos\theta|)\mp |d_i||\mathbf{v}|\right]+\dots\,,
\end{align}
where the final expression is solved for the velocity:
\begin{equation}
v^i=\frac{1}{m_{\psi}}(p^i\pm |\mathbf{d}||p^i\cos\theta|)\pm|d_i|\frac{|\mathbf{p}|}{m_{\psi}}+\dots\,.
\end{equation}
The Hamilton function can then be computed as
\begin{equation}
\label{eq:energy-momentum-d4-first}
\mathcal{H}^{\widehat{d}^{\,(4)}\pm}_1=m_{\psi}+\frac{\mathbf{p}^2}{2m_{\psi}}\pm |\mathbf{d}\cdot\mathbf{p}|\frac{|\mathbf{p}|}{m_{\psi}}+\dots\,.
\end{equation}
This corresponds to Eq.~(23) of \cite{Colladay:2012rv} at first order in Lorentz violation.

\subsubsection{Antisymmetric operator with nonvanishing timelike components}
\label{sec:d-space-case-2}

For the second case considered in \cite{Colladay:2012rv}, $d^{(4)}_{\mu\nu}$ is assumed to be antisymmetric. Furthermore,
the quantity $Y$ shall vanish. An important case that obeys these properties is an antisymmetric two-tensor $d^{(4)}_{\mu\nu}$ with
nonvanishing components only in the first row and column, respectively. Hence, the coefficients
$d^{(4)}_{0i}$ are taken to be nonvanishing again where $d^{(4)}_{i0}=-d^{(4)}_{0i}$. Additionally, we introduce
the same vector $\mathbf{d}$ as before:
\begin{equation}
d^{(4)}_{\mu\nu}=\begin{pmatrix}
0 & d^{(4)}_{01} & d^{(4)}_{02} & d^{(4)}_{03} \\
-d^{(4)}_{01} & 0 & 0 & 0 \\
-d^{(4)}_{02} & 0 & 0 & 0 \\
-d^{(4)}_{03} & 0 & 0 & 0 \\
\end{pmatrix}_{\mu\nu}\,,\quad \mathbf{d}\equiv\begin{pmatrix}
d^{(4)}_{01} \\
d^{(4)}_{02} \\
d^{(4)}_{03} \\
\end{pmatrix}\,.
\end{equation}
The Lagrange function can be found in Eq.~(42) of \cite{Colladay:2012rv}. In contrast to the previous Lagrangian of
\eqref{eq:lagrangian-d4-first}, the current one is written in covariant form:\footnote{The Lagrangian for
an antisymmetric $d^{(4)}_{\mu\nu}$ was also derived in Eq.~(16) of \cite{Russell:2015gwa}. Both results differ at second
order in Lorentz violation due to different signs before $X$. The plus sign seems to be the correct one \cite{Colladay:2016}.
This sign does not have any influence on our final results, though.}
\begin{subequations}
\label{eq:lagrangian-d4-2}
\begin{align}
L^{d^{(4)}\pm}_2&=-\frac{m_{\psi}}{1+2X}\left(\sqrt{(1+2X)u^2+D}\pm\sqrt{D}\right)\,, \\[2ex]
D&=\mathbf{d}^2(u^0)^2-(\mathbf{d}\cdot\mathbf{u})^2\,,\quad X=-\frac{\mathbf{d}^2}{2}\,.
\end{align}
\end{subequations}
Introducing the angle $\theta$ between the vector $\mathbf{d}$ and $\mathbf{v}$, we write the energy as follows:
\begin{align}
E^{d^{(4)}\pm}_2&=\frac{m_{\psi}}{1-\mathbf{d}^2}\left[\frac{1}{\sqrt{1+(\mathbf{d}^2-1)\mathbf{v}^2-(\mathbf{d}\cdot\mathbf{v})^2}}\pm\frac{\mathbf{d}^2}{\sqrt{\mathbf{d}^2-(\mathbf{d}\cdot\mathbf{v})^2}}\right] \notag \\
&=m_{\psi}\left(1+\frac{1}{2}\mathbf{v}^2\right)\pm|\mathbf{d}|m_{\psi}\left(1+\frac{1}{2}\mathbf{v}^2\cos^2\theta\right)+\dots\,.
\end{align}
The momentum-velocity correspondence is given by
\begin{subequations}
\begin{align}
p^i&=\frac{m_{\psi}}{1-\mathbf{d}^2}\left[\frac{(1-\mathbf{d}^2)v^i+d_i(\mathbf{d}\cdot\mathbf{v})}{\sqrt{1+(\mathbf{d}^2-1)\mathbf{v}^2-(\mathbf{d}\cdot\mathbf{v})^2}}\pm\frac{d_i(\mathbf{d}\cdot\mathbf{v})}{\sqrt{\mathbf{d}^2-(\mathbf{d}\cdot\mathbf{v})^2}}\right] \notag \\
&=m_{\psi}\left[v^i\pm d_i|\mathbf{v}|\cos\theta\right]+\dots\,, \\[2ex]
\label{eq:velocity-momentum-correspondence-second}
v^i&=\frac{p^i}{m_{\psi}}\mp d_i\frac{|\mathbf{p}|}{m_{\psi}}\cos\theta+\dots\,.
\end{align}
\end{subequations}
Here, the same issue arises that we encountered for the cases of the $b$ and $H$ coefficients considered in
\secref{sec:b-space-spacelike} and \secref{sec:H-space-case-1}, respectively.
Absolute-value bars around $\cos\theta$ have to be taken into account when solving the momentum-velocity correspondence for
the velocity. We eliminate those as we did before where for spin pointing up, the upper sign in \eqref{eq:velocity-momentum-correspondence-second}
is taken for $\theta\in [0,\pi/2]$ and the lower for $\theta\in (\pi/2,\pi]$. The result is then employed to obtain the
Hamilton function:
\begin{align}
\mathcal{H}^{\widehat{d}^{\,(4)}\pm}_2&=m_{\psi}+\frac{\mathbf{p}^2}{2m_{\psi}}\pm |\mathbf{d}|m_{\psi}\left(1-\frac{\mathbf{p}^2}{2m_{\psi}^2}\cos^2\theta\right)+\dots \notag \\
&=m_{\psi}+\frac{\mathbf{p}^2}{2m_{\psi}}\pm m_{\psi}\left(|\mathbf{d}|-\frac{1}{2m_{\psi}^2}\frac{(\mathbf{d}\cdot\mathbf{p})^2}{|\mathbf{d}|}\right)+\dots\,.
\end{align}
The latter is written in terms of the scalar product of $\mathbf{d}$ and $\mathbf{p}$ with higher-order contributions
in Lorentz violation neglected. Note the difference to the first case of \eqref{eq:energy-momentum-d4-first} that is
linear in the scalar product.

\subsubsection{Antisymmetric operator with nonvanishing spatial components}
\label{sec:d-space-case-3}

Another case with $Y=0$ can be constructed by choosing $d_{\mu\nu}^{(4)}$ to be antisymmetric with nonvanishing spatial
components only:
\begin{subequations}
\begin{align}
d^{(4)}_{\mu\nu}&=\begin{pmatrix}
0 & 0 & 0 & 0 \\
0 & 0 & d^{(4)}_{12} & d^{(4)}_{13} \\
0 & -d^{(4)}_{12} & 0 & d^{(4)}_{23} \\
0 & -d^{(4)}_{13} & -d^{(4)}_{23} & 0 \\
\end{pmatrix}_{\mu\nu}\,,\quad \mathbf{d}\equiv\begin{pmatrix}
d^{(4)}_{23} \\
-d^{(4)}_{13} \\
d^{(4)}_{12} \\
\end{pmatrix}\,, \displaybreak[0]\\[2ex]
D&=\mathbf{d}^2\mathbf{u}^2-(\mathbf{d}\cdot\mathbf{u})^2\,,\quad X=\frac{1}{2}\mathbf{d}^2\,.
\end{align}
\end{subequations}
The Lagrangian is taken from \eqref{eq:lagrangian-d4-2} where the parameters given above have to be inserted.
The energy can be computed as usual:
\begin{equation}
E^{\widehat{d}^{\,(4)}}_3=\frac{m_{\psi}}{\sqrt{1-\mathbf{v}^2+\mathbf{d}^2-(\mathbf{d}\cdot\mathbf{v})^2}}=m_{\psi}\left(1+\frac{1}{2}\mathbf{v}^2\right)+\dots\,.
\end{equation}
We observe that there is no first-order Lorentz-violating contribution as there are only quadratic terms in
$\mathbf{d}$. This is different from the particle momentum and velocity where $\mathbf{d}$ appears at linear order:
\begin{subequations}
\begin{align}
p^i&=\frac{m_{\psi}}{1+\mathbf{d}^2}\left[\frac{v^i+d_i(\mathbf{d}\cdot\mathbf{v})}{\sqrt{1-\mathbf{v}^2+\mathbf{d}^2-(\mathbf{d}\cdot\mathbf{v})^2}}\mp\frac{\mathbf{d}^2v^i-d_i(\mathbf{d}\cdot\mathbf{v})}{\sqrt{\mathbf{d}^2\mathbf{v}^2-(\mathbf{d}\cdot\mathbf{v})^2}}\right] \notag \\
&=m_{\psi}\left(v^i\mp|\mathbf{d}|\frac{v^i}{|\mathbf{v}|}\csc\theta\pm d_i\cot\theta\right)+\dots\,, \\[2ex]
v^i&=p^i\left(\frac{1}{m_{\psi}}\pm \frac{|\mathbf{d}|}{|\mathbf{p}|}\csc\theta\right)\mp d_i\cot\theta+\dots\,.
\end{align}
\end{subequations}
Note that both expressions are singular in $|\mathbf{p}|$ and $|\mathbf{v}|$, respectively. Furthermore, the plus
sign has to be taken for $p^1\geq 0$ and the minus sign for $p^1<0$. This behavior was
also observed for the timelike case of the $b$ coefficients in \secref{sec:b-space-timelike} and the timelike
sector of the $H$ coefficients, cf.~\secref{sec:H-space-case-2}. Replacing the velocity by the momentum in the
particle energy introduces Lorentz-violating terms into the Hamilton function:
\begin{equation}
\mathcal{H}^{\widehat{d}^{\,(4)}}_3=m_{\psi}+\frac{\mathbf{p}^2}{2m_{\psi}}\pm |\mathbf{d}||\mathbf{p}|\sin\theta+\dots=m_{\psi}+\frac{\mathbf{p}^2}{2m_{\psi}}\pm |\mathbf{d}\times\mathbf{p}|+\dots
\end{equation}
As usual, we neglect second-order Lorentz-violating contributions, which allows for introducing the cross product
between $\mathbf{d}$ and the three-momentum. In contrast to the momentum and velocity, the Hamilton function
does not have any singularities.

\subsection{Operator \texorpdfstring{$\boldsymbol{\widehat{g}^{(4)}}$}{g4-hat}}

Finally, we would like to consider the dimensionless observer tensor coefficients $g^{(4)}_{\mu\nu\varrho}$ that can be taken
as antisymmetric in the first two indices, cf.~\cite{Kostelecky:2000mm}. In \cite{Russell:2015gwa} some interesting cases were
studied. According to \cite{Fittante:2012ua}, the $g$ coefficients can be decomposed into an axial
part, a trace part, and a mixed-symmetry part. To do so, the axial vector $A^{\mu}$ and the trace vector $T_{\mu}$
are introduced, cf.~\cite{Russell:2015gwa}:
\begin{subequations}
\begin{align}
\label{eq:axial-vector}
A^{\alpha}&\equiv \frac{1}{6}g^{(4)}_{\sigma\kappa\tau}\varepsilon^{\sigma\kappa\tau\alpha}\,, \\[2ex]
\label{eq:trace-vector}
T_{\nu}&\equiv \frac{1}{3}g^{(4)\alpha}_{\nu\alpha}\,.
\end{align}
\end{subequations}

\subsubsection{Spacelike axial part}
\label{sec:g-space-case-1}

The first case of $g$ coefficients that will be investigated is included in the axial case. We take the following
choice of coefficients:
\begin{equation}
g^{(4)}_{023}\equiv g_1\equiv -g^{(4)}_{320}\,,\quad g^{(4)}_{013}\equiv -g_2\equiv -g^{(4)}_{310}\,,\quad g^{(4)}_{012}\equiv g_3\equiv -g^{(4)}_{210}\,,
\end{equation}
where coefficients with cyclic permutations of indices have the same value and anticyclic ones get an additional
minus sign. With this choice the timelike component of the axial vector $A^{\mu}$ vanishes and the spacelike
components take the values
\begin{subequations}
\begin{align}
A^1&\equiv\frac{1}{6}g^{(4)}_{\sigma\kappa\tau}\varepsilon^{\sigma\kappa\tau 1}= g_1\,, \\[2ex]
A^2&\equiv\frac{1}{6}g^{(4)}_{\sigma\kappa\tau}\varepsilon^{\sigma\kappa\tau 2}= g_2\,, \\[2ex]
A^3&\equiv\frac{1}{6}g^{(4)}_{\sigma\kappa\tau}\varepsilon^{\sigma\kappa\tau 3}= g_3\,.
\end{align}
\end{subequations}
Explicitly, the axial vector reads $A^{\mu}=(0,\mathbf{g})^{\mu}$ with $\mathbf{g}\equiv(g_1,g_2,g_3)$.
The Lagrangian that covers this case is given by Eq.~(30) in \cite{Russell:2015gwa}:
\begin{align}
L^{\widehat{g}^{(4)}\pm}_1&=-\frac{m_{\psi}}{1+A^2}\left(\sqrt{u^2+(A\cdot u)^2}\pm\sqrt{(A\cdot u)^2-A^2u^2}\right) \notag \\
&=-\frac{m_{\psi}}{1-\mathbf{g}^2}\left(\sqrt{u^2+(\mathbf{g}\cdot\mathbf{u})^2}\pm\sqrt{(\mathbf{g}\cdot\mathbf{u})^2+\mathbf{g}^2u^2}\right)\,.
\end{align}
Note that the second term in the Lagrangian is of bipartite form just as for the $b$ coefficients, cf.~\eqref{eq:lagrangian-b3}.
To compute the energy we introduce the angle $\theta$ between the vectors $\mathbf{g}$ and $\mathbf{v}$ in analogy to
the $b$, $H$, and $d$ cases. This leads to
\begin{align}
E^{\widehat{g}^{(4)}\pm}_1&=\frac{m_{\psi}}{1-\mathbf{g}^2}\left[\frac{1}{\sqrt{1-\mathbf{v}^2+(\mathbf{g}\cdot\mathbf{v})^2}}\pm\frac{\mathbf{g}^2}{\sqrt{(1-\mathbf{v}^2)\mathbf{g}^2+(\mathbf{g}\cdot\mathbf{v})^2}}\right] \notag \\
&=m_{\psi}\left(1+\frac{1}{2}\mathbf{v}^2\right)\pm |\mathbf{g}|m_{\psi}\left(1+\frac{1}{2}\mathbf{v}^2\sin^2\theta\right)+\dots\,.
\end{align}
The same procedure is applied to obtain the momentum:
\begin{align}
p^i&=\frac{m_{\psi}}{1-\mathbf{g}^2}\left[\frac{v^i-g_i(\mathbf{g}\cdot\mathbf{v})}{\sqrt{1-\mathbf{v}^2+(\mathbf{g}\cdot\mathbf{v})^2}}\mp\frac{g_i(\mathbf{g}\cdot\mathbf{v})-\mathbf{g}^2v^i}{\sqrt{(1-\mathbf{v}^2)\mathbf{g}^2+(\mathbf{g}\cdot\mathbf{v})^2}}\right] \notag \\
&=(1\pm|\mathbf{g}|)m_{\psi}v^i\mp g_im_{\psi}|\mathbf{v}|\cos\theta+\dots\,.
\end{align}
Solving the latter expression for the velocity results in
\begin{equation}
v^i=(1\mp|\mathbf{g}|)\frac{p^i}{m_{\psi}}\pm g_i\frac{|\mathbf{p}|}{m_{\psi}}\cos\theta+\dots\,.
\end{equation}
Finally, the Hamilton function reads
\begin{align}
\mathcal{H}^{\widehat{g}^{(4)}\pm}_1&=m_{\psi}+\frac{\mathbf{p}^2}{2m_{\psi}}\pm |\mathbf{g}|m_{\psi}\left(1-\frac{\mathbf{p}^2}{2m_{\psi}^2}\sin^2\theta\right)+\dots \notag \\
&=m_{\psi}+\frac{\mathbf{p}^2}{2m_{\psi}}\pm m_{\psi}\left(|\mathbf{g}|-\frac{1}{2m_{\psi}^2}\frac{(\mathbf{g}\times\mathbf{p})^2}{|\mathbf{g}|}\right)+\dots\,.
\end{align}
Here the cross product of $\mathbf{g}$ and $\mathbf{p}$ has been introduced as before with higher-order
terms in Lorentz violation neglected. With the particle being in spin-up state, the upper sign is valid for
$\theta\in [0,\pi/2]$ and the lower for $\theta\in (\pi/2,\pi]$, cf.~Secs.~\ref{sec:b-space-spacelike},
\ref{sec:H-space-case-1}, \ref{sec:d-space-case-2}. For spin pointing down, both signs must be switched.

\subsubsection{Partially antisymmetric tensor}
\label{sec:g-space-case-2}

The next case that shall be looked at was considered in \cite{Colladay:2012rv}. It is characterized by a choice of
coefficients of the form $g^{(4)}_{0ij}=\varepsilon_{ijk}g_k$, i.e., explicitly we have
\begin{subequations}
\begin{align}
g^{(4)}_{023}=g^{(4)}_{302}\equiv g_1\,,\quad g^{(4)}_{032}=g^{(4)}_{203}\equiv -g_1\,, \displaybreak[0]\\[2ex]
g^{(4)}_{031}=g^{(4)}_{103}\equiv g_2\,,\quad g^{(4)}_{013}=g^{(4)}_{301}\equiv -g_2\,, \displaybreak[0]\\[2ex]
g^{(4)}_{012}=g^{(4)}_{201}\equiv g_3\,,\quad g^{(4)}_{021}=g^{(4)}_{102}\equiv -g_3\,.
\end{align}
\begin{subequations}
Furthermore, we introduce $\mathbf{g}\equiv(g_1,g_2,g_3)$ to write everything in a convenient way. The choice made is not
totally antisymmetric in all three indices and, therefore, it differs from the sector considered in the previous subsection.
The Lagrangian is given by Eqs.~(29), (30) in \cite{Colladay:2012rv}:
\begin{align}
L^{\widehat{g}^{(4)}\pm}_2&=-m_{\psi}\sqrt{1-\frac{\mathbf{v}^2}{C_{g\pm}}}\,, \\[2ex]
C_{g\pm}&=\left(\sqrt{1-(\widehat{\mathbf{v}}\cdot\mathbf{g})^2}\pm |\widehat{\mathbf{v}}\times\mathbf{g}|\right)^2=\left(\sqrt{1-\mathbf{g}^2\cos^2\theta}\pm|\mathbf{g}|\sin\theta\right)^2\,,
\end{align}
\end{subequations}
and it has a structure similar to \eqref{eq:lagrangian-d4-first} for the $d$ coefficients, which was obtained
in the same paper. The current Lagrangian does not depend on $u^0$ in analogy to \eqref{eq:lagrangian-d4-first}
since it was also derived in proper-time parameterization. Therefore, a Legendre transformation according to
\eqref{eq:legendre-transform} must be carried out to compute the energy. With the angle $\theta$ between $\mathbf{g}$ and $\mathbf{v}$
the result reads
\begin{equation}
E^{\widehat{g}^{(4)}\pm}_2=\frac{m_{\psi}}{\sqrt{1-\mathbf{v}^2/C_{\pm}}}=m_{\psi}\left(1+\frac{1}{2}\mathbf{v}^2\right)\mp|\mathbf{g}|m_{\psi}\mathbf{v}^2\sin\theta+\dots
\end{equation}
The momentum is obtained similarly:
\end{subequations}
\begin{align}
p^i&=\frac{m_{\psi}}{C_{\pm}\sqrt{C_{\pm}-\mathbf{v}^2}}\left[v^i\sqrt{C_{\pm}}-(\mathbf{g}\cdot\mathbf{v})\left[(\mathbf{g}\cdot\mathbf{v})v^i-g_i\mathbf{v}^2\right]\right. \notag \\
&\phantom{{}={}\frac{m_{\psi}}{C_{\pm}\sqrt{C_{\pm}-\mathbf{v}^2}}\Big[}\times\left.\frac{1}{|\mathbf{v}|}\left(\frac{1}{\sqrt{\mathbf{v}^2-(\mathbf{g}\cdot\mathbf{v})^2}}\pm\frac{1}{\sqrt{\mathbf{g}^2\mathbf{v}^2-(\mathbf{g}\cdot\mathbf{v})^2}}\right)\right] \notag \\
&=m_{\psi}v^i\pm g_im_{\psi}|\mathbf{v}|\cot\theta\mp |\mathbf{g}|m_{\psi}v^i(\sin\theta+\csc\theta)+\dots\,.
\end{align}
Finally, the velocity
\begin{equation}
v^i=\frac{p^i}{m_{\psi}}\mp g_i\frac{|\mathbf{p}|}{m_{\psi}}\cot\theta\pm |\mathbf{g}|\frac{p^i}{m_{\psi}}(\sin\theta+\csc\theta)+\dots\,,
\end{equation}
is employed to calculate the Hamilton function:
\begin{equation}
\mathcal{H}^{\widehat{g}^{(4)}\pm}_2=m_{\psi}+\frac{\mathbf{p}^2}{2m_{\psi}}\pm|\mathbf{g}|\frac{\mathbf{p}^2}{m_{\psi}}\sin\theta+\dots=m_{\psi}+\frac{\mathbf{p}^2}{2m_{\psi}}\pm |\mathbf{g}\times\mathbf{p}|\frac{|\mathbf{p}|}{m_{\psi}}+\dots
\end{equation}
Contrary to the first case of $g$ coefficients considered, this result is linear in the magnitude of the cross
product between $\mathbf{g}$ and $\mathbf{p}$. Furthermore, it is equal to Eq.~(28) of \cite{Colladay:2012rv} taken at first
order in Lorentz violation. When the particle is in spin-up state the upper sign holds for $\theta\in [0,\pi/2]$ and the
lower for $\theta\in (\pi/2,\pi]$,
cf.~Secs.~\ref{sec:b-space-spacelike}, \ref{sec:H-space-case-1}, \ref{sec:d-space-case-2}, and \ref{sec:g-space-case-1}.

\subsubsection{Trace part}
\label{sec:g-space-case-3}

The final sector of $g$ coefficients to be considered is based on a nonvanishing trace vector of \eqref{eq:trace-vector}.
The choice of coefficients is as follows:
\begin{subequations}
\begin{align}
g^{(4)}_{011}=g^{(4)}_{022}=g^{(4)}_{033}&\equiv -g_0\,, \displaybreak[0]\\[2ex]
g^{(4)}_{100}=-g^{(4)}_{122}=-g^{(4)}_{133}&\equiv g_1\,, \displaybreak[0]\\[2ex]
g^{(4)}_{200}=-g^{(4)}_{211}=-g^{(4)}_{233}&\equiv g_2\,, \displaybreak[0]\\[2ex]
g^{(4)}_{300}=-g^{(4)}_{311}=-g^{(4)}_{322}&\equiv g_3\,,
\end{align}
\end{subequations}
where furthermore $g^{(4)}_{\nu\mu\varrho}=-g^{(4)}_{\mu\nu\varrho}$ to respect antisymmetry in the first two indices.
The trace vector then has the form $(T_{\mu})=(g_0,g_1,g_2,g_3)$. The corresponding Lagrangian can be found in Eq.~(29)
of~\cite{Russell:2015gwa}. Interestingly, this Lagrangian does not comprise any linear-order term in Lorentz violation,
but the corrections are of second order:
\begin{equation}
L^{\widehat{g}^{(4)}}_3=-\frac{m_{\psi}}{1+T^2}\sqrt{u^2+(T\cdot u)^2}=-m_{\psi}\sqrt{u^2}+\mathcal{O}(T^2)\,.
\end{equation}
Therefore, for this case the energy and momentum have second-order effects in Lorentz violation only. We will come back
to this point later.

\subsection{Concluding remarks}

The most interesting cases have been studied in the previous subsections. In principle, other sectors could be
investigated such as the traceless, diagonal choice
\begin{equation}
d^{(4)}_{\mu\nu}=d^{(4)}_{00}\mathrm{diag}\left(1,\frac{1}{3},\frac{1}{3},\frac{1}{3}\right)\,,
\end{equation}
for the $d$ coefficients. The latter has $X\neq 0$ and $Y=0$, which makes the Lagrangian of \eqref{eq:lagrangian-d4-2}
applicable resulting in isotropic expressions. However, those cases do not provide further insight, which is why they will
be skipped.

\section{First quantization and Hamilton operators}
\label{sec:first-quantization}

In the previous section we obtained the Hamilton functions $\mathcal{H}=\mathcal{H}(\mathbf{p})$ for a series of classical
Lagrangians associated to the minimal SME. These Hamilton functions depend on the spatial momentum $\mathbf{p}$ and the
Lorentz-violating controlling coefficients involved. For practical reasons, we present here a complete list of those:
\begin{subequations}
\label{eq:hamilton-functions-all}
\begin{align}
\mathcal{H}^{\widehat{a}^{(3)}}&\approx\alpha+a^{(3)}_0+\frac{\mathbf{a}\cdot\mathbf{p}}{m_{\psi}}\,, \displaybreak[0]\\[2ex]
\mathcal{H}^{\widehat{c}^{\,(4)}}&\approx(1-c^{(4)}_{00})\alpha-2(\mathbf{c}\cdot\mathbf{p})-\frac{1}{m_{\psi}}c^{(4)}_{ij}p^ip^j\,, \displaybreak[0]\\[2ex]
\mathcal{H}^{\widehat{e}^{\,(4)}}&\approx\alpha-(m_{\psi}e^{(4)}_0+\mathbf{e}\cdot\mathbf{p})\,, \displaybreak[0]\\[2ex]
\mathcal{H}^{\widehat{b}^{(3)}\pm}_1&\approx\alpha\pm\left(|\mathbf{b}|-\frac{1}{2m_{\psi}^2}\frac{(\mathbf{b}\times\mathbf{p})^2}{|\mathbf{b}|}\right)\,, \displaybreak[0]\\[2ex]
\mathcal{H}^{\widehat{b}^{(3)}\pm}_2&\approx\alpha\pm |b^{(3)}_0|\frac{|\mathbf{p}|}{m_{\psi}}\,, \displaybreak[0]\\[2ex]
\mathcal{H}^{\widehat{H}^{(3)}\pm}_1&\approx\alpha\mp\left(|\mathbf{h}|-\frac{1}{2m_{\psi}^2}\frac{(\mathbf{h}\cdot\mathbf{p})^2}{|\mathbf{h}|}\right)\,, \displaybreak[0]\\[2ex]
\mathcal{H}^{\widehat{H}^{(3)}\pm}_2&\approx\alpha\mp \frac{1}{m_{\psi}}|\mathbf{h}\times\mathbf{p}|\,, \displaybreak[0]\\[2ex]
\mathcal{H}^{\widehat{H}^{(3)}\pm}_3&\approx\alpha\mp \left[|\mathbf{h}|+\frac{1}{2m_{\psi}^2}\left(\frac{(\mathbf{h}\times\mathbf{p})^2}{|\mathbf{h}|}-\frac{(\mathbf{h}\cdot\mathbf{p})^2}{|\mathbf{h}|}\right)\right]\,, \displaybreak[0]\\[2ex]
\mathcal{H}^{\widehat{d}^{\,(4)}\pm}_1&\approx\alpha\pm |\mathbf{d}\cdot\mathbf{p}|\frac{|\mathbf{p}|}{m_{\psi}}\,, \displaybreak[0]\\[2ex]
\mathcal{H}^{\widehat{d}^{\,(4)}\pm}_2&\approx\alpha\pm m_{\psi}\left(|\mathbf{d}|-\frac{1}{2m_{\psi}^2}\frac{(\mathbf{d}\cdot\mathbf{p})^2}{|\mathbf{d}|}\right)\,, \displaybreak[0]\\[2ex]
\mathcal{H}^{\widehat{d}^{\,(4)}\pm}_3&\approx\alpha\pm |\mathbf{d}\times\mathbf{p}|\,, \displaybreak[0]\\[2ex]
\mathcal{H}^{\widehat{g}^{(4)}\pm}_1&\approx\alpha\pm m_{\psi}\left(|\mathbf{g}|-\frac{1}{2m_{\psi}^2}\frac{(\mathbf{g}\times\mathbf{p})^2}{|\mathbf{g}|}\right)\,, \displaybreak[0]\\[2ex]
\mathcal{H}^{\widehat{g}^{(4)}\pm}_2&\approx\alpha\pm |\mathbf{g}\times\mathbf{p}|\frac{|\mathbf{p}|}{m_{\psi}}\,,
\end{align}
with
\begin{equation}
\label{eq:definition-alpha}
\alpha\equiv m_{\psi}+\frac{\mathbf{p}^2}{2m_{\psi}}+\dots\,.
\end{equation}
\end{subequations}
These now serve as the base for first quantization in analogy to
standard quantum mechanics. Note that the Hamilton operator obtained from the minimal SME Lagrange density (at first order
in Lorentz violation and second order in the momentum) can be found in \cite{Kostelecky:1999mr,Kostelecky:1999zh,Yoder:2012ks}.
In the first two papers the result was derived for the first time and in the third reference the Hamilton operator is used
for computing SME-induced corrections to the energy levels of the hydrogen atom. In the latter reference, the symmetries of
the controlling coefficients were additionally taken into account to state the Hamiltonian. Our results will be compared to
theirs for consistency.

Since for the cases of the $a$, $c$, and $e$ coefficients the spin degeneracy is not broken
the Hamilton operators for these sectors can be obtained in a straightforward way. The classical momentum $\mathbf{p}$
is understood to be replaced by a quantum-mechanical operator $\widehat{\mathbf{p}}$ where just as in standard quantum
mechanics we have that
\begin{equation}
\widehat{\mathbf{p}}=\frac{1}{\mathrm{i}}\frac{\partial}{\partial\widehat{\mathbf{x}}}\,,\quad [\widehat{p}^{\,i},\widehat{x}^j]=-\mathrm{i}\delta_{ij}\,,
\end{equation}
with the position operator $\widehat{\mathbf{x}}$ and the Kronecker delta $\delta_{ij}$.
In spin space the Hamilton operators $H$ of the $a$, $c$, and $e$ coefficients are merely proportional to the $(2\times 2)$
identity matrix $\mathds{1}_2$. Therefore, for these sectors we obtain:
\begin{equation}
H^{\widehat{a}^{(3)}}=\mathcal{H}^{\widehat{a}^{(3)}}(\widehat{\mathbf{p}})\mathds{1}_2\,,\quad H^{\widehat{c}^{\,(4)}}=\mathcal{H}^{\widehat{c}^{(4)}}(\widehat{\mathbf{p}})\mathds{1}_2\,,\quad H^{\widehat{e}^{\,(4)}}=\mathcal{H}^{\widehat{e}^{(4)}}(\widehat{\mathbf{p}})\mathds{1}_2\,,
\end{equation}
where the Hamilton functions are taken from Eqs.~(\ref{eq:energy-momentum-a3}), (\ref{eq:energy-momentum-c4}), and
(\ref{eq:energy-momentum-e4}). Since spin degeneracy is broken for the $b$, $H$, $d$, and $g$ coefficients, it is more
difficult to obtain the quantum-mechanical Hamilton operators for those sectors. Just as before, the classical momenta
in the Hamilton functions have to be replaced by momentum operators. However, in spin space the Hamilton operators cannot
be proportional to the identity matrix but they are expected to involve the Pauli matrices
\begin{equation}
\sigma_1=\begin{pmatrix}
0 & 1 \\
1 & 0 \\
\end{pmatrix}\,,\quad \sigma_2=\begin{pmatrix}
0 & -\mathrm{i} \\
\mathrm{i} & 0 \\
\end{pmatrix}\,,\quad \sigma_3=\begin{pmatrix}
1 & 0 \\
0 & -1 \\
\end{pmatrix}\,.
\end{equation}
Therefore, for these Hamilton operators the following \textit{Ansatz} is reasonable:
\begin{equation}
\label{eq:hamilton-operator-ansatz}
H=\Xi\,\mathds{1}_2+B_i\sigma^i+D_{ij}\frac{p^i}{m_{\psi}}\sigma^j+F_{ijk}\frac{p^ip^j}{m_{\psi}^2}\sigma^k\,.
\end{equation}
Here $\Xi$ is a scalar in spin space. For the $a$, $c$, and $e$ coefficients this is the only nonvanishing contribution
and the real parameters $B_i$, $D_{ij}$, and $F_{ijk}$ vanish for these cases. The energy eigenvalues for a Hamilton
operator of the form proposed in \eqref{eq:hamilton-operator-ansatz} are given by:
\begin{align}
\label{eq:energy-eigenvalues-hamilton-operator}
E&=\Xi\pm \left(\mathbf{B}^2+2B_aD_{ia}\frac{p^i}{m_{\psi}}+\left(D_{ia}D_{ja}+2B_aF_{ija}\right)\frac{p^ip^j}{m_{\psi}^2}\right. \notag \\
&\phantom{{}={}\Xi\pm \Bigg)}\left.+\,2D_{ia}F_{jka}\frac{p^ip^jp^k}{m_{\psi}^3}+F_{ija}F_{kla}\frac{p^ip^jp^kp^l}{m_{\psi}^4}\right)^{1/2} \notag \\
&=\Xi\pm \left\{|\mathbf{B}|+\frac{1}{|\mathbf{B}|}\left[B_aD_{ia}\frac{p^i}{m_{\psi}}+\frac{1}{2}\left(D_{ia}D_{ja}+2B_aF_{ija}\right)\frac{p^ip^j}{m_{\psi}^2}\right]\right\}+\dots\,.
\end{align}
The result stated in the first line is exact where in the second line it has been expanded for $(p^i)^2\ll m_{\psi}^2$.
The latter expansion is only valid for $B_i\neq 0$. For some cases $B_i=0$ and $F_{ijk}=0$ resulting in
\begin{equation}
E\Big|_{\substack{B_i=0 \\ F_{ijk}=0}}=\Xi\pm\frac{1}{m_{\psi}}\sqrt{D_{ia}D_{ja}p^ip^j}\,,
\end{equation}
where for others $B_i=0$ and $D_{ij}=0$ leading to
\begin{equation}
E\Big|_{\substack{B_i=0 \\ D_{ij}=0}}=\Xi\pm\frac{1}{m_{\psi}^2}\sqrt{F_{ija}F_{kla}p^ip^jp^kp^l}\,.
\end{equation}
Comparing the latter expansions to the energies obtained in the previous section allows for computing the parameters
$B_i$, $D_{ij}$, and $F_{ijk}$. In general, for the $b$, $H$, $d$, and $g$ coefficients the scalar part $\Xi$ does
not involve any Lorentz-violating terms, but it holds that $\Xi=\alpha$ with $\alpha$ given in \eqref{eq:definition-alpha}.

It was found that the aforementioned Hamilton operator of \eqref{eq:hamilton-operator-ansatz} can be obtained with this
procedure in a straightforward way for almost all of the sectors considered. The results are summarized in \tabref{tab:parameters-hamilton-operators}
and they are consistent with \cite{Kostelecky:1999mr,Kostelecky:1999zh,Yoder:2012ks}. There is only one exception where the method seems to
be more challenging: the case of $\widehat{H}^{(3)}$ with $X=0$, $Y\neq 0$ discussed in \secref{sec:H-space-case-3}.
Inspecting the Hamilton function of \eqref{eq:energy-momentum-H3-second} reveals that both $B_i$ and $F_{ijk}$ must be
nonzero. From $B_i\neq 0$ one could then deduce that $D_{ij}=0$ because there is no term linear in the momentum.
However, a direct comparison to the latter references reveals that $B_aD_{ia}=0$ only but $D_{ij}\neq 0$:
\begin{table}[t]
\centering
\begin{tabular}{ccccc}
\toprule
SME sector & Subsection & $B_i$ & $D_{ij}$ & $F_{ijk}$ \\
\colrule
Spacelike $\widehat{b}^{(3)}$ & \ref{sec:b-space-spacelike} & $-b_i^{(3)}$ & $0$ & $\varepsilon_{jka}\varepsilon_{aip}b_p^{(3)}/2$ \\
Timelike $\widehat{b}^{(3)}$ & \ref{sec:b-space-timelike} & 0 & $-b^{(3)}_0\delta_{ij}$ & 0 \\
$\widehat{H}^{(3)}$ with $X\neq 0$, $Y=0$ (timelike) & \ref{sec:H-space-case-1} & $\varepsilon_{iab}H^{(3)}_{ab}/2$ & 0 & $-\varepsilon_{iab}H^{(3)}_{ab}\delta_{jk}/4$ \\
$\widehat{H}^{(3)}$ with $X\neq0$, $Y=0$ (spacelike) & \ref{sec:H-space-case-2} & 0 & $\varepsilon_{ija}H^{(3)}_{a0}$ & 0 \\
$\widehat{H}^{(3)}$ with $X=0$, $Y\neq 0$ & \ref{sec:H-space-case-3} & $\varepsilon_{iab}H^{(3)}_{ab}/2$ & $\varepsilon_{ija}H^{(3)}_{a0}$ & $-\varepsilon_{iab}H^{(3)}_{ab}\delta_{jk}/4$ \\
Nonsymmetric $\widehat{d}^{\,(4)}$ & \ref{sec:d-space-case-1} & 0 & 0 & $m_{\psi}d^{(4)}_{0i}\delta_{jk}$ \\
Antisymmetric $\widehat{d}^{\,(4)}$ (timelike) & \ref{sec:d-space-case-2} & $-m_{\psi}d_{0i}^{(4)}$ & 0 & $m_{\psi}d^{(4)}_{0i}\delta_{jk}/2$ \\
Antisymmetric $\widehat{d}^{\,(4)}$ (spacelike) & \ref{sec:d-space-case-3} & 0 & $m_{\psi}d^{(4)}_{ji}$ & 0 \\
Axial $\widehat{g}^{(4)}$ & \ref{sec:g-space-case-1} & $-m_{\psi}\varepsilon_{iab}g^{(4)}_{ab0}/2$ & 0 & $-m_{\psi}\varepsilon_{jka}(2g^{(4)}_{a0i}+g^{(4)}_{ai0})/2$ \\
Partially antisymmetric $\widehat{g}^{(4)}$ & \ref{sec:g-space-case-2} & 0 & 0 & $-m_{\psi}\varepsilon_{jka}(2g^{(4)}_{a0i}+g^{(4)}_{ai0})/2$ \\
Trace $\widehat{g}^{(4)}$ & \ref{sec:g-space-case-3} & 0 & 0 & 0 \\
\botrule
\end{tabular}
\caption{Parameters of the Hamilton operator given by \eqref{eq:hamilton-operator-ansatz} for the $b$, $H$, $d$, and $g$
sectors considered in the previous section.}
\label{tab:parameters-hamilton-operators}
\end{table}
\begin{align}
B_qD_{iq}&=\frac{1}{2}\varepsilon_{qab}H^{(3)}_{ab} \varepsilon_{iqc}H^{(3)}_{c0}=-\frac{1}{2}(\delta_{ai}\delta_{bc}-\delta_{ac}\delta_{bi})H^{(3)}_{ab}H^{(3)}_{c0}=\frac{1}{2}(H^{(3)}_{ci}H^{(3)}_{c0}-H^{(3)}_{ic}H^{(3)}_{c0}) \notag \\
&=H_{ci}^{(3)}H^{(3)}_{c0}
=0_i\,.
\end{align}
Finally, the third sector of $g$ coefficients considered in \secref{sec:g-space-case-3} did not deliver any contribution
at first order in Lorentz violation. This is in accordance with \cite{Kostelecky:1999mr,Kostelecky:1999zh,Yoder:2012ks}
taking into account that
\begin{equation}
\frac{1}{2}\varepsilon^{abk}g^{(4)}_{abi}+\varepsilon^{ika}g^{(4)}_{a00}=0_{ik}\,,
\end{equation}
for this choice of coefficients.

\section{Limit of Lagrangians and observations}
\label{sec:limit-of-lagrangians}

Some further interesting results can be obtained directly from the Lagrangians that we considered in this article. So far,
we have just computed the expanded Hamilton functions. Now we again work in proper-time parameterization and perform expansions
of the Lagrangians at first order in Lorentz violation and at second order in the three-velocity:
\begin{subequations}
\label{eq:limit-lagrangians}
\begin{align}
L^{\widehat{a}^{(3)}}&\approx-\beta-(a^{(3)}_0+\mathbf{a}\cdot\mathbf{v})\,, \displaybreak[0]\\[2ex]
L^{\widehat{c}^{\,(4)}}&\approx-\beta+c^{(4)}_{00}m_{\psi}\left(1+\frac{1}{2}\mathbf{v}^2\right)+2m_{\psi}(\mathbf{c}\cdot\mathbf{v})+m_{\psi}c^{(4)}_{ij}v^iv^j\,, \displaybreak[0]\\[2ex]
L^{\widehat{e}^{(4)}}&\approx-\beta+m_{\psi}(e^{(4)}_0+\mathbf{e}\cdot\mathbf{v})\,, \displaybreak[0]\\[2ex]
L^{\widehat{b}^{(3)}\pm}_1&\approx-\beta\mp |\mathbf{b}|\left(1-\frac{1}{2}\mathbf{v}^2\sin^2\theta\right)=-\beta\mp\left(|\mathbf{b}|-\frac{1}{2}\frac{(\mathbf{b}\times\mathbf{v})^2}{|\mathbf{b}|}\right)\,, \displaybreak[0]\\[2ex]
L^{\widehat{b}^{(3)}\pm}_2&\approx-\beta\mp |b^{(3)}_0||\mathbf{v}|\,, \displaybreak[0]\\[2ex]
L^{\widehat{H}^{(3)}\pm}_1&\approx-\beta\pm |\mathbf{h}|\left(1-\frac{1}{2}\mathbf{v}^2\cos^2\theta\right)=-\beta\pm\left(|\mathbf{h}|-\frac{1}{2}\frac{(\mathbf{h}\cdot\mathbf{v})^2}{|\mathbf{h}|}\right)\,, \displaybreak[0]\\[2ex]
L^{\widehat{H}^{(3)}\pm}_2&\approx-\beta\pm |\mathbf{h}|v\sin\theta=-\beta\pm |\mathbf{h}\times\mathbf{v}|\,, \displaybreak[0]\\[2ex]
L^{\widehat{H}^{(3)}\pm}_3&\approx-\beta\pm |\mathbf{h}|\left(1-\frac{1}{2}\mathbf{v}^2\cos(2\theta)\right)=-\beta\pm \left[|\mathbf{h}|+\frac{1}{2}\left(\frac{(\mathbf{h}\times\mathbf{v})^2}{|\mathbf{h}|}-\frac{(\mathbf{h}\cdot\mathbf{v})^2}{|\mathbf{h}|}\right)\right]\,, \displaybreak[0]\\[2ex]
L^{\widehat{d}^{\,(4)}\pm}_1&\approx-\beta\mp |\mathbf{d}|m_{\psi}\mathbf{v}^2|\cos\theta|=-\beta\mp m_{\psi}|\mathbf{d}\cdot\mathbf{v}||\mathbf{v}|\,, \displaybreak[0]\\[2ex]
L^{\widehat{d}^{\,(4)}\pm}_2&\approx-\beta\mp |\mathbf{d}|m_{\psi}\left(1-\frac{1}{2}\mathbf{v}^2\cos^2\theta\right)=-\beta\mp m_{\psi}\left(|\mathbf{d}|-\frac{1}{2}\frac{(\mathbf{d}\cdot\mathbf{v})^2}{|\mathbf{d}|}\right)\,, \displaybreak[0]\\[2ex]
L^{\widehat{d}^{\,(4)}\pm}_3&\approx-\beta\mp |\mathbf{d}|m_{\psi}|\mathbf{v}|\sin\theta=-\beta\mp m_{\psi}|\mathbf{d}\times\mathbf{v}|\,, \displaybreak[0]\\[2ex]
L^{\widehat{g}^{(4)}\pm}_1&\approx-\beta\mp |\mathbf{g}|m_{\psi}\left(1-\frac{1}{2}\mathbf{v}^2\sin^2\theta\right)=-\beta\mp m_{\psi}\left(|\mathbf{g}|-\frac{1}{2}\frac{(\mathbf{g}\times\mathbf{v})^2}{|\mathbf{g}|}\right)\,, \displaybreak[0]\\[2ex]
L^{\widehat{g}^{(4)}\pm}_2&\approx-\beta\mp |\mathbf{g}|m_{\psi}\mathbf{v}^2\sin\theta=-\beta\mp m_{\psi}|\mathbf{g}\times\mathbf{v}||\mathbf{v}|\,,
\end{align}
where
\begin{equation}
\beta\equiv m_{\psi}\left(1-\frac{1}{2}\mathbf{v}^2\right)+\dots\,.
\end{equation}
\end{subequations}
Inspecting these expansions closely leads to a number of interesting observations. First, except of the term $m_{\psi}c^{(4)}_{ij}v^iv^j$
in the Lagrangian of the $c$ coefficients all first-order expansions merely involve vector magnitudes, scalar products and cross
products, i.e., the structure of the expansions is very limited. Second, in almost all expansions there is either a scalar product
or a cross product. The only exception is $L^{\widehat{H}^{(3)}\pm}_3$ that involves both. From these observations we can deduce a
number of interesting properties of classical Lagrangians for the minimal SME fermion sector.

\begin{itemize}

\item[1)] It seems that only Lagrangians whose expansions involve either a scalar product or a cross product can be derived in a
closed and simple form. Note that $L^{\widehat{H}^{(3)}\pm}_3$ is presumably the most complicated result found in the minimal SME
since it follows as a zero of the involved polynomial given in Eq.~(18) of \cite{Kostelecky:2010hs}. No other Lagrangian has been
obtained with a more complicated expansion than the ones stated above.
\item[2)] Comparing the expanded Lagrangians to the Hamilton functions of Eqs.~(\ref{eq:hamilton-functions-all}) we discover a great
deal of similarities. It seems that there is the connection
\begin{equation}
L=\alpha-\beta-\mathcal{H}|_{p^i=m_{\psi}v^i}+\dots=m_{\psi}\mathbf{v}^2-\mathcal{H}|_{p^i=m_{\psi}v^i}+\dots\,,
\end{equation}
for a Lagrangian $L$ and Hamilton function $\mathcal{H}$ at first order in Lorentz violation and second order in the
velocity and momentum, respectively. This forms the basis of a proposition that shall be formulated below.

\end{itemize}

\begin{conjecture}
\label{con:connection-lagrangian-hamiltonian}
Let $\mathcal{H}$ be the Hamilton function that corresponds to the modified dispersion relation of a massive fermion in the
minimal Standard-Model Extension. Under the assumption that the Hamilton function can be expressed
in terms of scalar products and vector products of the particle velocity and preferred spatial directions as well as in terms of
the magnitudes of these vectors, a classical Lagrangian $L$ is related to $\mathcal{H}$ as follows:
\begin{equation}
\label{eq:lagrangian-new-method}
L=m_{\psi}\mathbf{v}^2-\mathcal{H}|_{p^i=m_{\psi}v^i}+\mathcal{O}[k_x(v^i)^3]\,.
\end{equation}
Here $m_{\psi}$ is the particle mass, $\mathbf{v}$ its velocity, $\mathbf{p}$ the three-momentum, and $k_x$ are generic
Lorentz-violating coefficients.
\end{conjecture}

Apparently, \eqref{eq:lagrangian-new-method} is just a Legendre transformation. However, note that for it to be a proper Legendre
transformation the replacement rules for the momentum should comprise Lorentz-violating coefficients, cf.~all examples that we
considered in \secref{sec:classical-hamilton-functions}. A general proof of the validity of \eqref{eq:lagrangian-new-method}
is sketched in \appref{sec:sketch-proof-conjecture} based on the quantum mechanical transition amplitude
between two states of different particle position. The proposition allows for deriving (at least approximated) classical Lagrangians
even for the most involved dispersion relations of the minimal SME fermion sector. Such results could be useful for nonrelativistic
calculations. In what follows, we will consider two of these complicated cases.

\subsection{Applying the proposition}

\subsubsection{Example of $d$ coefficients}

To demonstrate the previously proposed method to deriving approximations for classical Lagrangians for involved cases of the
SME fermion sector we consider the following particular choice of $d$ coefficients. This example bears similarities to the choice
of $\widehat{H}^{(3)}$ that we made in \secref{sec:H-space-case-2}. Therefore, we introduce a timelike preferred direction
$\xi_{\mu}$ and a spacelike one $\zeta_{\mu}$. The (traceless) coefficient matrix $d^{(4)}_{\mu\nu}$ is then constructed as follows:
\begin{subequations}
\begin{align}
d^{(4)}_{\mu\nu}&=|\mathbf{d}|\xi_{\mu}\xi_{\nu}+\frac{1}{|\mathbf{d}|}\zeta_{\mu}\zeta_{\nu}+\xi_{\mu}\zeta_{\nu}-\zeta_{\mu}\xi_{\nu}-\varepsilon_{\mu\nu}^{\phantom{\mu\nu}\varrho\sigma}\xi_{\varrho}\zeta_{\sigma}\,, \\[2ex]
\xi_{\mu}&=\begin{pmatrix}
1 \\
\mathbf{0} \\
\end{pmatrix}\,,\quad \zeta_{\mu}=\begin{pmatrix}
0 \\
\mathbf{d} \\
\end{pmatrix}\,,\quad \mathbf{d}=\begin{pmatrix}
d \\
0 \\
0 \\
\end{pmatrix}\,.
\end{align}
\end{subequations}
For this choice $X=\mathbf{d}^2/2$ and $Y=-\mathbf{d}^2$ whereby attempts to obtaining Lagrangians for sectors with both $X\neq 0$
and $Y\neq 0$ have not been successful so far. The reason lies in the structure of the dispersion relation, which is very complicated
for such cases. The dispersion relation follows from the determinant of the modified Dirac equation, cf.~Eqs.~(2), (4), (6),
and (7) of \cite{Kostelecky:2013rta}. Using the vectors stated above it can be cast into the form
\begin{subequations}
\begin{align}
0&=(p^2-m_{\psi}^2)^2-2\left[2\mathbf{d}^2p_0^4-4|\mathbf{d}|(\zeta\cdot p)p_0^3+5(\zeta\cdot p)^2p_0^2+\mathbf{d}^2(p^2+m_{\psi}^2)T\right. \notag \displaybreak[0]\\
&\phantom{{}={}}\left.+\,4|\mathbf{d}|(\zeta\cdot p)p_0(\mathbf{p}^2-m_{\psi}^2)+(\zeta\cdot p)^2(m_{\psi}^2-T)\right] \notag \displaybreak[0]\\
&\phantom{{}={}}+\left[\frac{1}{2}\mathbf{d}^2(\mathbf{p}^2+T)-4p_0|\mathbf{d}|(\zeta\cdot p)\right]^2\,, \displaybreak[0]\\[2ex]
T&=(\xi\cdot p)^2-p^2-2\frac{(\zeta\cdot p)^2}{\mathbf{d}^2}\,.
\end{align}
\end{subequations}
The result involves terms with odd powers of $(\zeta\cdot p)$ that violate parity invariance. Since the $d^{(4)}_{\mu\nu}$ chosen
has nonvanishing $d_{00}^{(4)}$ and $d_{ij}^{(4)}$ coefficients, this is in accordance with Table P31 of~\cite{Kostelecky:2008ts}.
Although the choice of coefficients is fairly simple, the dispersion relation is already quite complicated.
Its zeros with respect to $p_0$ deliver the Hamilton function. The exact result involves third roots and is not illuminating,
which is why it will be skipped. However, an expansion at leading order in Lorentz violation and at second order in the momenta
is short enough to be given:
\begin{equation}
\label{eq:hamilton-function-first-example}
\mathcal{H}=p_0=m_{\psi}+\frac{\mathbf{p}^2}{2m_{\psi}}\pm \left\{|\mathbf{d}|m_{\psi}-2(\mathbf{d}\cdot\mathbf{p})+\frac{1}{2|\mathbf{d}|m_{\psi}}\left[2\mathbf{d}^2\mathbf{p}^2-3(\mathbf{d}\cdot\mathbf{p})^2\right]\right\}+\dots\,.
\end{equation}
Comparing to the Hamilton functions of Eqs.~(\ref{eq:hamilton-functions-all}) it becomes clear that the structure of the latter result
is more involved than that of the previous ones. After all, it comprises a scalar product between the spatial direction and the
three-momentum where its square occurs as well. Equation (\ref{eq:lagrangian-new-method}) allows for deriving the Lagrangian at this
level of approximation:
\begin{align}
L&=-\beta\mp |\mathbf{d}|m_{\psi}\left[1-2v^1-\frac{1}{2}(v^1)^2+(v^2)^2+(v^3)^2\right]+\dots \notag \\
&=-\beta\mp m_{\psi}\left\{|\mathbf{d}|(1+\mathbf{v}^2)-2(\mathbf{d}\cdot\mathbf{v})-\frac{3}{2}\frac{(\mathbf{d}\cdot\mathbf{v})^2}{|\mathbf{d}|}\right\}+\dots\,.
\end{align}
With the group velocity components
\begin{equation}
v^i=\frac{\partial p_0}{\partial p^i}=(1+2|\mathbf{d}|)\frac{p^i}{m_{\psi}}-d_i\left(2+\frac{1}{m_{\psi}}\frac{3(\mathbf{d}\cdot\mathbf{p})}{|\mathbf{d}|}\right)\,,
\end{equation}
and $L=-(\mathcal{H}-\mathbf{p}\cdot\mathbf{v})$ it can be demonstrated that the Lagrange function found fulfills
Eqs.~(\ref{eq:set-equations-lagrangians}) at first order in Lorentz violation and at second order in the velocities.
Comparing the Lagrangian to the expansions of all known results, Eqs.~(\ref{eq:limit-lagrangians}), emphasizes its
complicated structure. Since both the dispersion relation and the Lagrangian has been expressed by observer rotation
invariants an observer rotation can be performed to arrive at a $\mathbf{d}$ pointing along an arbitrary direction.

\subsubsection{Example of $g$ coefficients}

The Lagrangians for $g$ coefficients obtained in \cite{Russell:2015gwa} are valid for the axial and trace part of
$g^{(4)}_{\mu\nu\varrho}$ as indicated before. For the remaining ``mixed'' coefficients no Lagrangian has been
found until now. In this subsection such a case shall be considered. We take the choice $g^{(4)}_{102}=g^{(4)}_{201}\equiv g\equiv -g^{(4)}_{012}=-g^{(4)}_{021}$
where the antisymmetry of $g^{(4)}_{\mu\nu\varrho}$ in its first two indices is manifest. Furthermore, for this case
both the axial and trace vector of Eqs.~(\ref{eq:axial-vector}), (\ref{eq:trace-vector}) vanishes. Therefore, the
nonvanishing coefficients must belong to the mixed part of $g^{(4)}_{\mu\nu\varrho}$. The total tensor can be
expressed via a timelike direction $\xi_{\mu}$ and two spacelike directions $\zeta_{\mu}$, $\psi_{\mu}$ as follows:
\begin{subequations}
\begin{align}
g^{(4)}_{\mu\nu\varrho}&=\frac{1}{|\mathbf{g}_1|}\left[\zeta_{\mu}\xi_{\nu}\psi_{\varrho}+\psi_{\mu}\xi_{\nu}\zeta_{\varrho}-\left(\xi_{\mu}\zeta_{\nu}\psi_{\varrho}+\xi_{\mu}\psi_{\nu}\zeta_{\varrho}\right)\right]\,, \\[2ex]
(\xi_{\mu})&=\begin{pmatrix}
1 \\
\mathbf{0} \\
\end{pmatrix}\,,\quad (\zeta_{\mu})=\begin{pmatrix}
0 \\
\mathbf{g}_1 \\
\end{pmatrix}\,,\quad (\psi_{\mu})=\begin{pmatrix}
0 \\
\mathbf{g}_2 \\
\end{pmatrix}\,,\quad \mathbf{g}_1=\begin{pmatrix}
g \\
0 \\
0 \\
\end{pmatrix}\,,\quad \mathbf{g}_2=\begin{pmatrix}
0 \\
g \\
0 \\
\end{pmatrix}\,,
\end{align}
with $g>0$.
From the modified Dirac equation of \cite{Kostelecky:2013rta} we again obtain the exact dispersion relation. It can be expressed
via the preferred directions:
\end{subequations}
\begin{align}
0&=(p^2-m_{\psi}^2)^2+[(\zeta\cdot p)^2+(\psi\cdot p)^2]\left[(\zeta\cdot p)^2+(\psi\cdot p)^2-2(p_0^2+\mathbf{p}^2-m_{\psi}^2)\right] \notag \\
&\phantom{{}={}}+\frac{16}{\mathbf{g}_1^2}(\zeta\cdot p)^2(\psi\cdot p)^2\,.
\end{align}
The Hamilton function at first order in the controlling coefficients and at second order in the momentum reads
\begin{subequations}
\label{eq:hamilton-function-second-example}
\begin{align}
\mathcal{H}=p_0&=m_{\psi}+\frac{\mathbf{p}^2}{2m_{\psi}}\pm\frac{1}{m_{\psi}}\Upsilon+\dots\,, \displaybreak[0]\\[2ex]
\Upsilon&=\sqrt{\left[(\mathbf{g}_1\cdot\mathbf{p})^2+(\mathbf{g}_2\cdot\mathbf{p})^2\right]\mathbf{p}^2-\frac{4}{\mathbf{g}_1^2}(\mathbf{g}_1\cdot\mathbf{p})^2(\mathbf{g}_2\cdot\mathbf{p})^2}\,.
\end{align}
\end{subequations}
The result comprises only even powers of the momentum, which is why there is no parity violation. That is again in accordance with
Table P31 of \cite{Kostelecky:2008ts}, which says that parity is conserved for $g$ coefficients with a single timelike index.
In comparison to all Hamilton functions considered in this paper, the latter has the most complicated structure. It involves various scalar
products between the momentum and the preferred directions under a square root function. It is very probable that the exact
Lagrangian cannot be obtained for this case due to its complexity. We expect that it is either impossible to solve
Eqs.~(\ref{eq:set-equations-lagrangians}) at all or that the solution is too complicated to serve any purpose in practice. However,
via the proposition previously formulated we can at least obtain the result at first order in Lorentz violation and second order in
the velocity:
\begin{align}
L&=-\beta\mp m_{\psi}\sqrt{\left[(\mathbf{g}_1\cdot\mathbf{v})^2+(\mathbf{g}_2\cdot\mathbf{v})^2\right]\mathbf{v}^2-\frac{4}{\mathbf{g}_1^2}(\mathbf{g}_1\cdot\mathbf{v})^2(\mathbf{g}_2\cdot\mathbf{v})^2}+\dots\,.
\end{align}
Using the group velocity components
\begin{subequations}
\begin{align}
v^1&=\frac{p^1}{m_{\psi}}\left(1+\frac{\mathbf{g}_1^2\mathbf{p}^2+(\mathbf{g}_1\cdot\mathbf{p})^2-3(\mathbf{g}_2\cdot\mathbf{p})^2}{\Upsilon}\right)\,, \\[2ex]
v^2&=\frac{p^2}{m_{\psi}}\left(1+\frac{\mathbf{g}_1^2\mathbf{p}^2+(\mathbf{g}_2\cdot\mathbf{p})^2-3(\mathbf{g}_1\cdot\mathbf{p})^2}{\Upsilon}\right)\,, \\[2ex]
v^3&=\frac{p^3}{m_{\psi}}\left(1+\frac{(\mathbf{g}_1\cdot\mathbf{p})^2+(\mathbf{g}_2\cdot\mathbf{p})^2}{\Upsilon}\right)\,,
\end{align}
\end{subequations}
it was checked that the Lagrangian found obeys Eqs.~(\ref{eq:set-equations-lagrangians}) at the level of approximation used. Its structure
is still complicated enough, since it has quartic polynomials of the velocity under a square root. With an observer rotation the
Lagrange function can be generalized to $\mathbf{g}_1$ and $\mathbf{g}_2$ pointing along arbitrary directions where $|\mathbf{g}_1|=|\mathbf{g}_2|$.

\section{Conclusions}
\label{sec:conclusions}

In this paper we studied several aspects of already known classical Lagrangians for the minimal fermion sector of the SME. Taking
this collection of Lagrangians as a starting point and hiding our knowledge of the SME at a first place, we tried to perform first
quantization. This should be expected to lead to the low-energy quantum-mechanical Hamilton operator following from the SME. The
Hamilton operator was obtained consistently for all cases considered where the procedure was challenging for a single case only.

From these investigations, an interesting bonus result emerged that we had not expected at the beginning of the analysis. The
structure of the leading-order expansion for any known Lagrangian was found to be quite simple. These leading-order terms involve
either a first or a second power of a scalar product or a vector product between the velocity and a preferred spatial direction.
This led us to the suspicion that all minimal Lagrangians that have not been found so far must have a more complicated structure,
which is comprised of, e.g., combinations of scalar and vector products. Additionally, it was observed that all leading-order approximations
are related to their corresponding classical Hamilton functions by a very simple transformation. That result was formulated as a
proposition and proven to be valid universally. For two complicated cases of $d$ and $g$ coefficients it enabled us to derive
Lagrangians at first order in Lorentz violation and at second order in the velocity.
The proposition provides us with an easy way to tackle even the most complicated sectors of the minimal SME.


\section{Acknowledgments}

It is a pleasure to thank V.~A.~Kosteleck\'{y} for helpful comments and the suggestion of proving Proposition~\ref{con:connection-lagrangian-hamiltonian}
using the path integral method. Furthermore, the author is indebted to D.~Colladay for useful discussions. This work was partially
funded by the Brazilian foundation FAPEMA.

\newpage
\begin{appendix}
\numberwithin{equation}{section}

\section{Sketch of proof for Proposition \ref{con:connection-lagrangian-hamiltonian}}
\label{sec:sketch-proof-conjecture}
\setcounter{equation}{0}

The proof for Proposition \ref{con:connection-lagrangian-hamiltonian} presented here is based on the quantum-mechanical transition
amplitude $Z$ for the propagation of a particle between two states at different coordinates $\mathbf{x}_0$, $\mathbf{x}$ and times
$0$, $T$. The transition amplitude can be obtained as the matrix element including the two states and the Hamilton operator of the
free particle subject to Lorentz violation. Inserting a complete set of momentum states allows for expressing the matrix element in
terms of the classical Hamilton function $\mathcal{H}$:
\begin{align}
\label{eq:transition-amplitude}
Z&=\left\langle \mathbf{x},T\left|\exp\left(-\frac{\mathrm{i}}{\hbar}H(\widehat{\mathbf{p}})T\right)\right|\mathbf{x}_0,0\right\rangle=\int \mathrm{d}^3p\,\left\langle \mathbf{x},T\left|\exp\left(-\frac{\mathrm{i}}{\hbar}H(\widehat{\mathbf{p}})T\right)\right|\mathbf{p}\right\rangle\langle \mathbf{p}|\mathbf{x}_0,0\rangle \notag \\
&=\frac{1}{2\pi\hbar} \int \mathrm{d}^3p\,\exp\left(-\frac{\mathrm{i}}{\hbar}\mathcal{H}(\mathbf{p})T+\frac{\mathrm{i}}{\hbar}\mathbf{p}\cdot(\mathbf{x}_0-\mathbf{x})\right)\,.
\end{align}
\begin{subequations}
The result for the transition amplitude is known to involve the free-particle action $S$:
\begin{equation}
\label{eq:transition-amplitude-via-action}
Z=N\exp\left(\frac{\mathrm{i}}{\hbar}S\right)=N\exp\left(\frac{\mathrm{i}}{\hbar}\int_0^T \mathrm{d}t\,L\right)\,,
\end{equation}
where $L$ is the Lagrangian of the particle and $N$ a prefactor. In general, the transition amplitude can be computed via the path integral
method. However, for the free particle it is simpler (and possible) to perform the computation via basic quantum mechanical methods just
as indicated above. Starting from the Hamilton function $\mathcal{H}$, we will compute $Z$ according to the second line of
\eqref{eq:transition-amplitude}. Subsequently, we will check whether the result corresponds to \eqref{eq:transition-amplitude-via-action}
with the Lagrangian obtained from $\mathcal{H}$ via Proposition \ref{con:connection-lagrangian-hamiltonian}. If the proposition is
valid there should be no contradiction.

We will cover the three possible generic cases for how Lorentz violation can appear in the Lagrangian at first order when it is additionally
restricted to squares in the velocity components. The first possibility is a contribution $\delta$ with $|\delta|\ll m_{\psi}$ that does not depend on
velocity. The Hamilton function then has the form $\mathcal{H}\supset \mathcal{H}^{(1)}=p^2/(2m_{\psi})+m_{\psi}-\delta$ where according to the
conjecture we obtain the Lagrangian $L\supset L^{(1)}=m_{\psi}v^2/2-m_{\psi}+\delta$. Now the transition amplitude for this case will be
computed and it suffices to work in one dimension. The integral is Gaussian and its result is taken from \eqref{eq:gauss-integral-1}:
\begin{align}
Z^{(1)}&=\frac{1}{2\pi\hbar} \int_{-\infty}^{\infty} \mathrm{d}p\,\exp\left[-\frac{\mathrm{i}}{\hbar}\left(\frac{p^2}{2m}+m_{\psi}-\delta\right)T+\frac{\mathrm{i}}{\hbar}p(x_0-x)\right] \notag \\
&=\frac{1}{2\pi\hbar}\sqrt{\frac{\pi}{\mathrm{i}T/(2m_{\psi}\hbar)}}\exp\left[-\frac{1}{\hbar^2}\frac{(x_0-x)^2}{4\mathrm{i}T/(2m_{\psi}\hbar)}-\frac{\mathrm{i}}{\hbar}(m_{\psi}-\delta)T\right] \notag \\
&=\sqrt{\frac{m_{\psi}}{2\pi\hbar \mathrm{i}T}}\exp\left\{\frac{\mathrm{i}}{\hbar}\left[m_{\psi}\frac{(x_0-x)^2}{2T}-(m_{\psi}-\delta)T\right]\right\}\,.
\end{align}
The complex prefactor is not important for us but the argument in the exponential function, which should correspond to the free-particle
action $S$ multiplied by $\mathrm{i}/\hbar$, cf.~\eqref{eq:transition-amplitude-via-action}. To check this we compute the action
from the Lagrangian proposed:
\begin{equation}
S\supset S^{(1)}=\int_0^T \mathrm{d}t\,L^{(1)}=\int_0^T \mathrm{d}t\,\left[\frac{1}{2}m_{\psi}v^2-(m_{\psi}-\delta)\right]=m_{\psi}\frac{(x_0-x)^2}{2T}-(m_{\psi}-\delta)T\,.
\end{equation}
Hence, there is no contradiction.
The second case includes Lorentz-violating terms that are linear in the velocity. The Hamilton function is proposed to be of the form
$\mathcal{H}\supset \mathcal{H}^{(2)}=\mathbf{p}^2/(2m_{\psi})+m_{\psi}-\varepsilon_ip^i/m_{\psi}$ and taking into account the
conjecture, the Lagrangian reads $L\supset L^{(2)}=m_{\psi}\mathbf{v}^2/2-m_{\psi}+\varepsilon_iv^i$. Here $\boldsymbol{\varepsilon}$
is a vector comprising Lorentz-violating coefficients where $|\varepsilon_i|\ll m_{\psi}$. The transition amplitude for
this second case is now a bit more involved to calculate due to the vectorial structure in the argument of the exponential
function. Fortunately, there is a general result for such a Gaussian integral given in \eqref{eq:gauss-integral-2}. Hence, we obtain
\begin{align}
Z^{(2)}&=\frac{1}{2\pi\hbar}\int \mathrm{d}^3p\,\exp\left[-\frac{\mathrm{i}}{\hbar}\left(\frac{\mathbf{p}^2}{2m_{\psi}}+m_{\psi}\right)T+\frac{\mathrm{i}}{\hbar}\left(x_0-x+\frac{T}{m_{\psi}}\varepsilon\right)_ip^i\right] \notag \\
&=\frac{\sqrt{\mathrm{i}}}{2\pi\hbar}\left(\frac{2\pi\hbar m_{\psi}}{T}\right)^{3/2}\exp\left[-\frac{1}{\hbar^2}\frac{1}{2\mathrm{i}T/(m_{\psi}\hbar)}\left(\mathbf{x}_0-\mathbf{x}+\frac{T}{m_{\psi}}\boldsymbol{\varepsilon}\right)^2-\frac{\mathrm{i}}{\hbar}m_{\psi}T\right] \notag \\
&=\sqrt{2\pi\hbar\mathrm{i}}\left(\frac{m_{\psi}}{T}\right)^{3/2}\exp\left\{\frac{\mathrm{i}}{\hbar}\left[m_{\psi}\frac{(\mathbf{x}_0-\mathbf{x})^2}{2T}-m_{\psi}T+\boldsymbol{\varepsilon}\cdot(\mathbf{x}-\mathbf{x}_0)\right]\right\}+\mathcal{O}(\varepsilon_i^2)\,.
\end{align}
The action based on the Lagrangian above can be computed as follows:
\begin{align}
S\supset S^{(2)}&=\int_0^T \mathrm{d}t\,L^{(2)}=\int_0^T \mathrm{d}t\,\left(\frac{1}{2}m_{\psi}\mathbf{v}^2-m_{\psi}+\varepsilon_iv^i\right) \notag \\
&=m_{\psi}\frac{(\mathbf{x}_0-\mathbf{x})^2}{2T}-m_{\psi}T+\boldsymbol{\varepsilon}\cdot(\mathbf{x}_0-\mathbf{x})\,.
\end{align}
This result is enclosed by the square brackets in the exponential function of $Z^{(2)}$ demonstrating consistency. Last but not least,
for the third case we consider a Lorentz-violating contribution that is quadratic in the velocity components. In general, the Hamilton
function can be written as $\mathcal{H}\supset \mathcal{H}^{(3)}=(\mathds{1}_3-\eta)_{ij}p^ip^j/(2m_{\psi})+m_{\psi}$ where $\mathds{1}_3$
is the unit matrix in three dimensions, i.e., we have the standard term plus perturbations comprised in the matrix $\eta$.
The components of $\eta$ are assumed to be much smaller than 1. Based on the conjecture, the Lagrangian would be
$L\supset L^{(3)}=(\mathds{1}_3+\eta)_{ij}m_{\psi}v^iv^j/2-m_{\psi}$. The transition amplitude is another Gaussian integral that can be
calculated from \eqref{eq:gauss-integral-2}:
\begin{align}
Z^{(3)}&=\frac{1}{2\pi\hbar}\int \mathrm{d}^3p\,\exp\left[-\frac{\mathrm{i}}{\hbar}\left(\frac{1}{2m_{\psi}}(\mathds{1}_3-\eta)_{ij}p^ip^j+m_{\psi}\right)T+\frac{\mathrm{i}}{\hbar}(x_0-x)_ip^i\right] \notag \\
&=\frac{\sqrt{\mathrm{i}}}{2\pi\hbar}\frac{[2\pi\hbar m_{\psi}/T]^{3/2}}{\sqrt{\det(\mathds{1}_3-\eta)}}\exp\left[-\frac{1}{\hbar^2}\frac{[(\mathds{1}_3-\eta)^{-1}]^{ij}(x_0-x)_i(x_0-x)_j}{2\mathrm{i}T/(m_{\psi}\hbar)}-\frac{\mathrm{i}}{\hbar}m_{\psi}T\right] \notag \\
&=\sqrt{\frac{2\pi\hbar\mathrm{i}}{\det(\mathds{1}_3-\eta)}}\left(\frac{m_{\psi}}{T}\right)^{3/2} \notag \\
&\phantom{{}={}}\times\exp\left\{\frac{\mathrm{i}}{\hbar}\left[m_{\psi}\frac{(\mathds{1}_3+\eta)_{ij}(x_0-x)^i(x_0-x)^j}{2T}-m_{\psi}T\right]\right\}+\mathcal{O}(\eta_{ij}^2)\,.
\end{align}
Here we used that the inverse matrix of $\mathds{1}_3-\eta$ is given by $\mathds{1}_3+\eta$ at first order in Lorentz violation, since
$(\mathds{1}_3-\eta)(\mathds{1}_3+\eta)=\mathds{1}_3+\mathcal{O}(\eta_{ij}^2)$. Now we compare this result to the free-particle action
that is obtained by direct calculation:
\begin{align}
S\supset S^{(3)}&=\int_0^T \mathrm{d}t\,L^{(3)}=\int_0^T \mathrm{d}t\,\left[\frac{1}{2}m_{\psi}(\mathds{1}_3+\eta)_{ij}v^iv^j-m_{\psi}\right] \notag \\
&=m_{\psi}\frac{(\mathds{1}_3+\eta)_{ij}(x_0-x)^i(x_0-x)^j}{2T}-m_{\psi}T\,.
\end{align}
Hence, the action is comprised in the argument of the exponential function in $Z^{(3)}$. The only caveat here is that the matrix $R$
in \eqref{eq:gauss-integral-2} must be symmetric, i.e., the matrix $\eta$ comprising Lorentz-violating coefficients should be symmetric
as well. At first order in Lorentz violation this is the case for the $c$ coefficients. As long as the structure of the Hamilton function
is solely composed of scalar products and vector products (cf.,~e.g., the Hamilton function of \eqref{eq:hamilton-function-first-example})
it holds as well:
\begin{subequations}
\begin{align}
(\mathbf{b}\cdot\mathbf{p})^2&=M^{(1)}_{ij}p^ip^j\,,\quad |\mathbf{b}\times\mathbf{p}|^2=M^{(2)}_{ij}p^ip^j\,, \\[2ex]
M^{(1)}&=\begin{pmatrix}
b_1^2 & b_1b_2 & b_1b_3 \\
b_1b_2 & b_2^2 & b_2b_3 \\
b_1b_3 & b_2b_3 & b_3^2 \\
\end{pmatrix}\,,\quad M^{(2)}=\mathbf{b}^2\mathds{1}_3-M^{(1)}\,.
\end{align}
\end{subequations}
Note that a more complicated form of the Hamilton function such as in \eqref{eq:hamilton-function-second-example} can be simplified
by introducing angles between the preferred directions and the velocity. Let $\theta_1$ be the angle between $\mathbf{g}_1$,
$\mathbf{v}$ and $\theta_2$ the angle between $\mathbf{g}_2$, $\mathbf{v}$ in the latter example. The Lorentz-violating term $\Upsilon$
in \eqref{eq:hamilton-function-second-example} is then
\begin{equation}
\Upsilon=|\mathbf{g}_1|\mathbf{p}^2\sqrt{\cos^2\theta_1+\cos^2\theta_2-4\cos^2\theta_1\cos^2\theta_2}\,,
\end{equation}
which is definitely of the form assumed in $\mathcal{H}^{(3)}$ above.
Hence, all the structures of possible Lorentz-violating terms, which according to the assumptions can occur at first order in Lorentz
violation and at second order in the velocity, have been covered. This completes the sketch of the proof of Proposition
\ref{con:connection-lagrangian-hamiltonian}.

\subsection{Useful Gaussian integrals}

There are two Gaussian integrals that are very helpful in the context of quantum-mechanical transition amplitudes. The first
(cf.~Eq.~(3.323.2) in \cite{Gradshteyn:2007}) is a merely one-dimensional integral and the second a multi-dimensional one:
\begin{align}
\label{eq:gauss-integral-1}
\int_{-\infty}^{\infty} \mathrm{d}q\,\exp(-aq^2+bq+c)&=\sqrt{\frac{\pi}{a}}\exp\left(\frac{b^2}{4a}+c\right)\,, \\[2ex]
\label{eq:gauss-integral-2}
\int \mathrm{d}^nq\,\exp\left(-\frac{1}{2}R_{ij}q^iq^j+S_iq^i\right)&=\frac{(2\pi)^{n/2}}{\sqrt{\det(R)}}\exp\left(\frac{1}{2}(R^{-1})^{ij}S_iS_j\right)\,,
\end{align}
\end{subequations}
where for the first integral $a$, $b$, $c\in \mathbb{C}$. In the second, $n\geq 1$ and $R_{ij}$ is assumed to be symmetric, which
also makes it invertible. The components $R_{ij}$ and $S_i$ can be complex numbers. Note that in the mathematics literature,
$\mathrm{Re}(a^2)>0$ and a positive definite matrix $R$ is assumed for reasons of convergence. We will not employ this restriction.
In principle, convergence can be enforced by adding a suitable infinitesimal contribution to the argument of the exponential
functions, which is set to zero in the final result. However, for the cases studied above the integrals can just be
computed without any convergence problems occurring.

\end{appendix}

\newpage


\end{document}